\documentclass{aa} 
\usepackage[varg]{txfonts}
\usepackage{graphicx}
\usepackage{bm}
\usepackage{color}
\usepackage{natbib}
\usepackage{hyperref}
\hypersetup{colorlinks=true,linkcolor=blue,citecolor=blue,urlcolor=blue}
\usepackage[toc,page]{appendix}
\usepackage{ulem}

\begin{document}


\title{Testing the validity of the ray-tracing code GYOTO}
\author{M. Grould \and T. Paumard \and G. Perrin}
\institute{LESIA, Observatoire de Paris, CNRS, Universit\'e Pierre  Marie Curie, Universit\'e Paris Diderot, 5 place Jules Janssen, 92190 Meudon, France} 
\date{Accepted 2016 April 22}


\abstract 
{In the next few years, the near-infrared interferometer GRAVITY will be able to observe the Galactic center. Astrometric data will be obtained with an anticipated accuracy of 10 $\mu$as. To analyze these future data, we have developed a code called GYOTO to compute orbits and ray-trace images.} 
{We want to assess the validity and accuracy of GYOTO in a variety of contexts, in particular for stellar astrometry in the Galactic center. 
Furthermore, we want to tackle and complete a study made on the astrometric displacements that are due to lensing effects of a star of the central parsec with GYOTO.}
{We first validate GYOTO in the weak-deflection limit (WDL) by studying primary caustics and primary critical curves obtained for a Kerr black hole. We compare GYOTO results to available analytical approximations and estimate GYOTO errors using an intrinsic estimator.  
In the strong-deflection limit (SDL), we choose to compare null geodesics computed by GYOTO and the ray-tracing code named Geokerr.
Finally, we use GYOTO to estimate the apparent astrometric displacements of a star for different angles from Sagittarius A* (Sgr A*).}
{In the WDL, we find a good coherence between GYOTO results and approximations. The maximal difference is around $10^{-5} \mu$as. 
Our intrinsic estimator finds a conservative uncertainty estimate also around $10^{-5} \mu$as. 
In the SDL, both ray-tracing codes find the same photon's coordinates with a maximal difference of about $10^{-3} \mu$as.
The shift of a star located behind the plane of sky containing Sgr A* is consistent with the current study.
In addition, the effect of lensing on any star in this plane of sky is a radial shift by 5 $\mu$as, independent of the distance from Sgr A* up to a very large distance.}
{We have demonstrated that GYOTO is accurate to a very high level, orders of magnitude better than the GRAVITY requirements. GYOTO is also valid in weak- and strong-deflection regimes and for very long integrations. At the astrometric precision that GRAVITY is aiming for, lensing effects must always be taken into account when fitting stellar orbits in the central parsec of the Galaxy.}
\maketitle


\section{Introduction}

GYOTO\footnote{freely available at the URL \url{http://GYOTO.obspm.fr}} (General relativitY OrbiT of Observatoire de Paris) is a ray-tracing code developed by \cite{2011CQGra..28v5011V}.
It integrates null and timelike geodesics in any analytical or numerical metrics. GYOTO can compute images and spectra for a variety of astrophysical objects, such as moving stars or accretion disks, around a Kerr black hole. 
The code's ability to take numerical metrics into account allows it to compute images or trajectories of stars orbiting exotic objects such as a boson star \citep{2014PhRvD..90b4068G, 2015arXiv151004170V}.

The main motivation for the development of GYOTO was to interpret the data to be obtained with the second-generation very long baseline telescope interferometry (VLTI) instrument GRAVITY \citep{2011Msngr.143...16E}. This instrument  observes stars and flares orbiting Sgr A*. It  probes spacetime near the central object with an expected astrometric accuracy of 10 $\mu$as. In a preliminary work, \cite{2014MNRAS.441.3477V} used GYOTO to show that GRAVITY is  capable of distinguishing an ejected blob from alternative flare models.

The stellar orbits measured by GRAVITY is affected by several effects such as periastron shift and Lense-Thirring effects \citep{2008ApJ...674L..25W, 2010PhRvD..81f2002M}. In addition, the individual astrometric measurements are affected by relativistic effects: time delay and lensing \citep{2012ApJ...753...56B}.

All these effects need to be considered in an apparent orbit model that will be fitted to the GRAVITY data, enabling the nature of Sgr A* to be constrained. Since the goal of GRAVITY is to deliver astrometry at an accuracy of 10~$\mu$as, models need to be more accurate than this value so as not to limit the accuracy of final results. We therefore aim for a model with an astrometric accuracy of 1 $\mu$as. In this paper, we study the accuracy of GYOTO to determine whether this tool can be used as a foundation for a future apparent orbit model to fit the GRAVITY data. Using the star images computed by GYOTO, it will be possible to get the apparent position of the star. However, the accuracy of this position will depend on the precision of the photon trajectories. Null geodesics need to be properly computed by the integrator implemented in GYOTO to take into account the correct bending effect. Besides, because of the 2'' field-of-view of GRAVITY, a wide range of distances between stars and Sgr A* will be possible. GYOTO has never been used with this type of  a configuration, so we need to ensure that geodesics are well computed. 

We first focus on the Einstein ring formation, more precisely we study primary caustics and primary critical curves considering a Kerr black hole.
The aims are both to compare our numerical results with the analytical study on primary caustics and primary critical curves performed by Sereno and De Luca in 2008 \citep{2008PhRvD..78b3008S}, and to check whether the numerical error is sufficiently low, which means inferior or equal to 1 $\mu$as. 
The comparison between GYOTO and approximations is a validation of GYOTO in the weak-deflection limit (WDL), however we also have to check if this ray-tracing code is valid in the strong-deflection limit (SDL). To do so, we choose to compare null geodesics computed in GYOTO and with another code named Geokerr\footnote{freely available at the URL 
 \url{http://www.astro.washington.edu/users/agol/geokerr/}} \citep{2009ApJ...696.1616D}. Before starting the validation and test of GYOTO, we discuss the different integrators implemented in it to choose the most appropriate one. To do so, we consider a Schwarzschild black hole and compare GYOTO with one of the approximations developed by \cite{2008PhRvD..78b3008S}.
 
In this paper, we also investigate lensing effects on the apparent position of a star to compare the astrometric displacements found in \cite{2012ApJ...753...56B}. We demonstrate that the effect cannot be neglected even when the star is in front of the plane of the sky containing Sgr~A*. We also discuss the impact of neglecting the lensing effects on the estimation of the orbital parameters and on the detection of other effects affecting the orbits (see Section~\ref{sec:five}).\\

The paper is organized as follows. In Section~\ref{sec:two}, we introduce our notations and define primary caustics and primary critical curves in the Schwarzschild and Kerr cases. We present the different analytical approximations we choose for our study. In Section~\ref{sec:three}, we discuss the validity of the different integrators implemented in GYOTO. Section~\ref{sec:four} is devoted to the comparison of the GYOTO results with those obtained by other methods: analytical formulae of primary caustics and primary critical curves (see Section~\ref{ssec:WDL}) and Geokerr (\ref{ssec:SDL}). Methods implemented to determine the different parameters with GYOTO are explained in Appendix~\ref{app:A}. A discussion on lensing effects is presented in Section~\ref{sec:five}. Concluding remarks are given in Section~\ref{sec:six}.


\section{The Einstein ring formation: focus on primary caustics and primary critical curves}
\label{sec:two}

\begin{figure}[!h]
\begin{center}
        \includegraphics[scale=0.5]{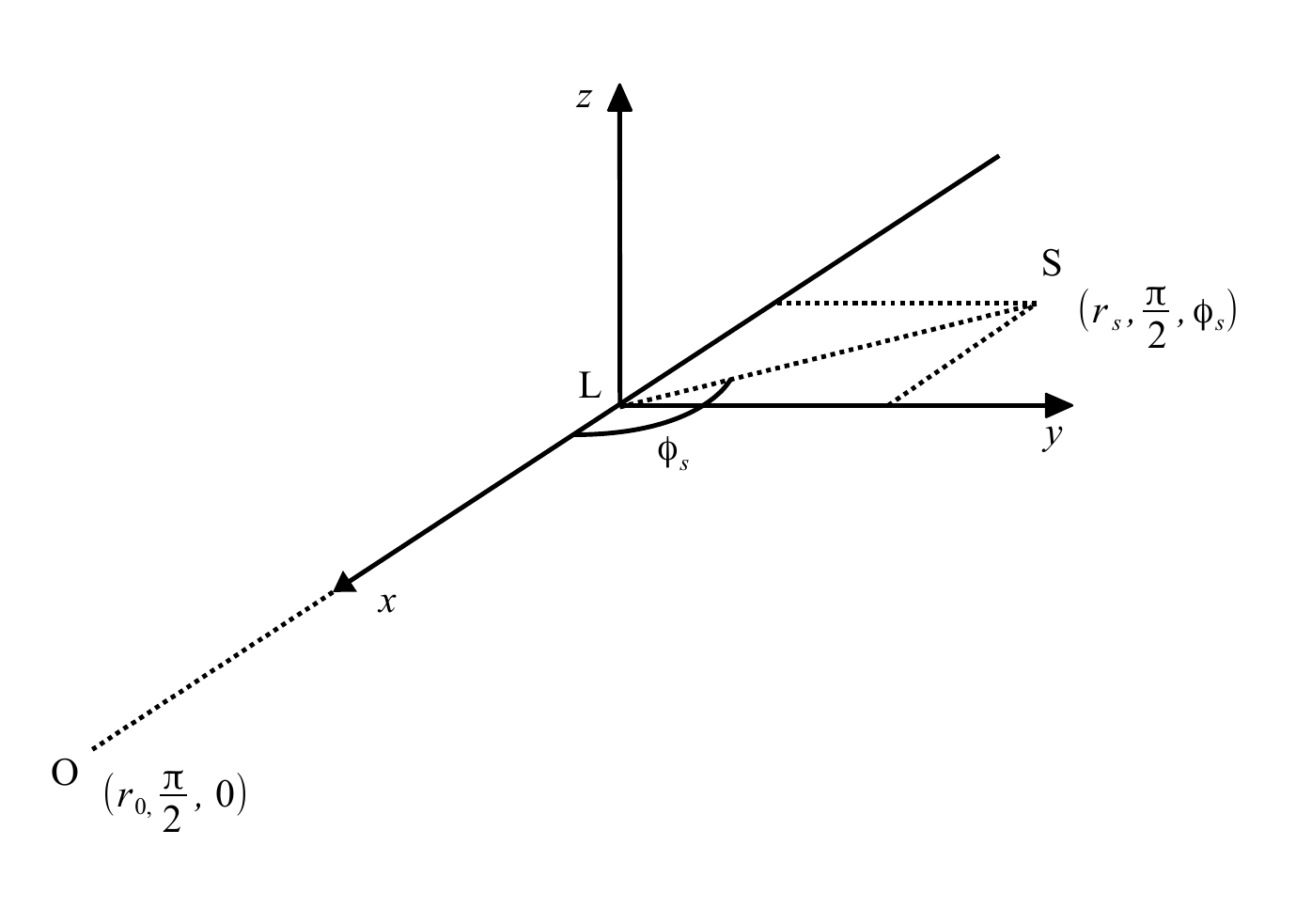}
        \caption{Lensing configuration: O, S, and L denote the observer, the source, and the lens, respectively.}
                \label{fig:axes}
\end{center}
\end{figure}

To understand how the Einstein ring is formed, we recall the basics of gravitational lensing using a Schwarzschild lens. 
Then, we focus on the Einstein ring obtained with a Kerr black hole. In both cases we consider a static observer. The geometry for lensing configuration is shown in Fig.~\ref{fig:axes}. 
The spin axis coincides with the z-axis. Spherical coordinates of the observer and the source, relative to the lens L, are noted $(r_0,\vartheta_0, \phi_0)$ and $(r_s,\vartheta_s,\phi_s)$, respectively. Without loss of generality, we choose to work in the equatorial plane of the black hole: $\vartheta_0=\pi/2$, $\phi_0 = 0$ and $\vartheta_s=\pi/2$. This yields $(x_0,0,0)$ for the observer and $(x_s,y_s,0)$ for the source. We note $M$ the lens mass and $a$ the spin of the black hole ranging from 0 (Schwarzschild black hole) to 1 (extremal Kerr black hole). From the point of view of the observer, the black hole rotates from the left to the right. In this paper, we use two different units for the distance: parsecs and geometrical units. This last unit is equal to $GM/c^{2}$ with $G$ the Newton's constant and $c$ the speed of light, but we will consider $G=c=1$ and note it $M$.


\subsection{Schwarzschild lens}

\begin{figure}[!h]
        \centering
        \includegraphics[scale=0.7]{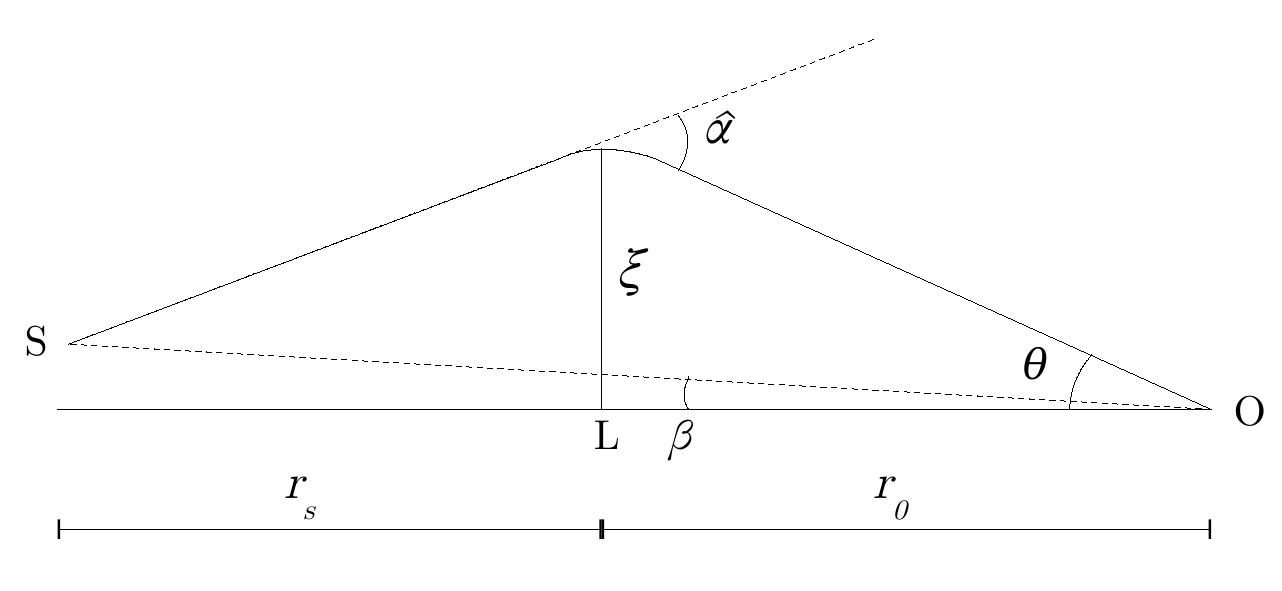}
        \caption{Spatial projection of a Schwarzschild lensing situation: S corresponds to the source, L to the lens and O to the observer.}
        \label{fig:lensconfig}
\end{figure}

Using the notation of Fig. \ref{fig:lensconfig}, we can write the lens equation \citep{1992grle.book.....S}
\begin{equation} 
        \bm{\beta} = \bm{\theta} - \frac{r_{s}}{(r_{s}+r_0)}\bm{\hat{\alpha}(\xi)},
\end{equation}
with $\bm{\beta}$ the unlensed angular position of the source, $\bm{\theta}$ the lensed angular position of the source equal to $\bm{\xi}/r_0$ and $\bm{\hat{\alpha}}$ the deflection angle depending on the impact parameter $
\bm{\xi}$.
The latter angle is given by
\begin{equation}
        \bm{\hat{\alpha}(\xi)} = \frac{2R_S}{\bm{\xi}},
\end{equation}
with $R_S=2GM/c^2$ the Schwarzschild radius. 
We can rewrite the lens equation as
\begin{equation}
        \bm{\theta}^2 - \bm{\beta}\bm{\theta} - \alpha_0^2 = 0,
\end{equation} 
with 
\begin{equation}
\alpha_0 = \sqrt{2R_S\frac{r_s}{r_0(r_s+r_0)}},
\end{equation} 
which corresponds to the Einstein angle.
The magnification of the source in the lens plane is a function of the lensed and unlensed angles as 
\begin{equation}
         A = \bigg| det \frac{\partial\bm{\beta}}{\partial\bm{\theta}}\bigg|^{-1}.
\end{equation}
$A$ is infinite when $det \left( \partial\bm{\beta}/\partial\bm{\theta} \right)=0$. In the source plane, these positions are called caustic points. For a Schwarzschild lens, the caustic is a line behind the lens starting from it and extending toward infinity \citep{1994ApJ...421...46R}. If the source lies on the caustic line then $\beta=0$. Thus, the solution of the lens equation is $\theta = \alpha_0$. A circle called a critical curve is formed in the lens plane with a radius of $\alpha_0$. Considering the source as a star, the observer sees the well-known Einstein ring. The radius of the ring corresponds to the critical curve radius so we get $\alpha_0=\theta_E$ with $\theta_E$ the Einstein ring radius. If the star does not lie on the caustic, the observer will see two images called primary and secondary images. These images are formed by lensing effects. 
Light rays are deviated because of the curvature of spacetime by the black hole. At the caustic points, the lensed images merge into the Einstein ring.
\begin{figure}[!t]
        \centering
        \includegraphics[scale=0.6]{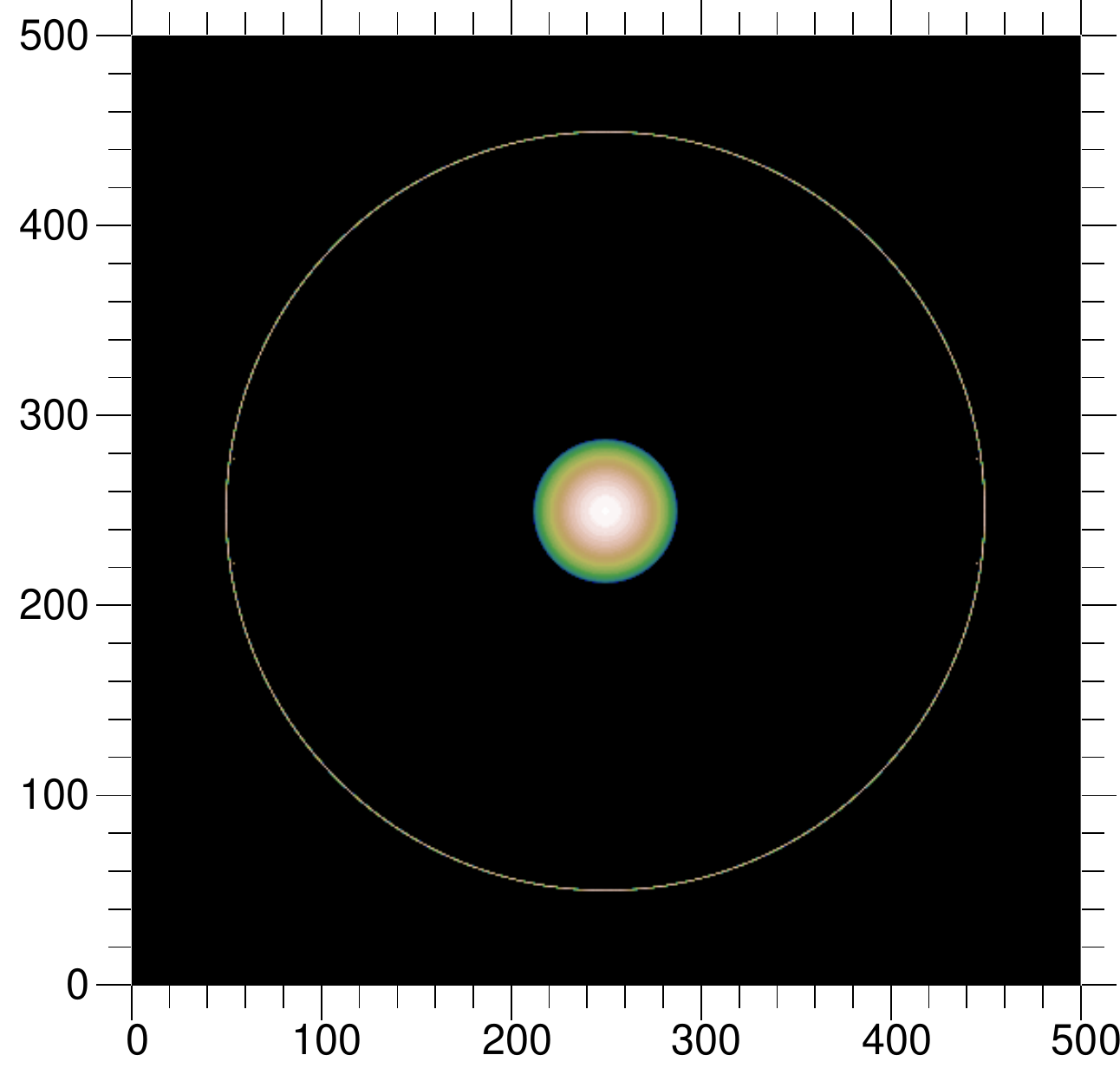}
        \caption{GYOTO image of a fixed star, with a radius of $1M$, located at 6$M$ in front of a Schwarzschild black hole. The observer is at 8 kpc from the black hole. The star lies on the second-order caustic line: the biggest image corresponds to the primary image of the star and the ring to the critical curve formed thanks to the second-order caustic. Axes are labeled in $\mu$as and the field of view of the image is equal to 70 $\mu$as.}
        \label{fig:ER2}
\end{figure}

Several studies have been performed on caustics and critical curves with Schwarzschild black holes. \cite{1994ApJ...421...46R} showed the existence of several orders of caustics. The caustic line we previously discussed is the first-order (also known as primary) caustic. This caustic is generated by light rays, which do not wind around the black hole. Higher orders correspond to photons winding one or several times around the black hole. \cite{2008PhRvD..78f3014B} showed that these caustics are also lines but caustics of even order start from the black hole and extend to the observer. It is also possible to form an Einstein ring if the star lies on these caustics of even order. But it is harder to detect as the largest fraction of the flux is concentrated in the primary image of the star (see Fig. \ref{fig:ER2}).


\subsection{Kerr black hole lens}
\label{ssec:KBHl}

\begin{figure}[!t]
\begin{center}
        \includegraphics[scale=0.6]{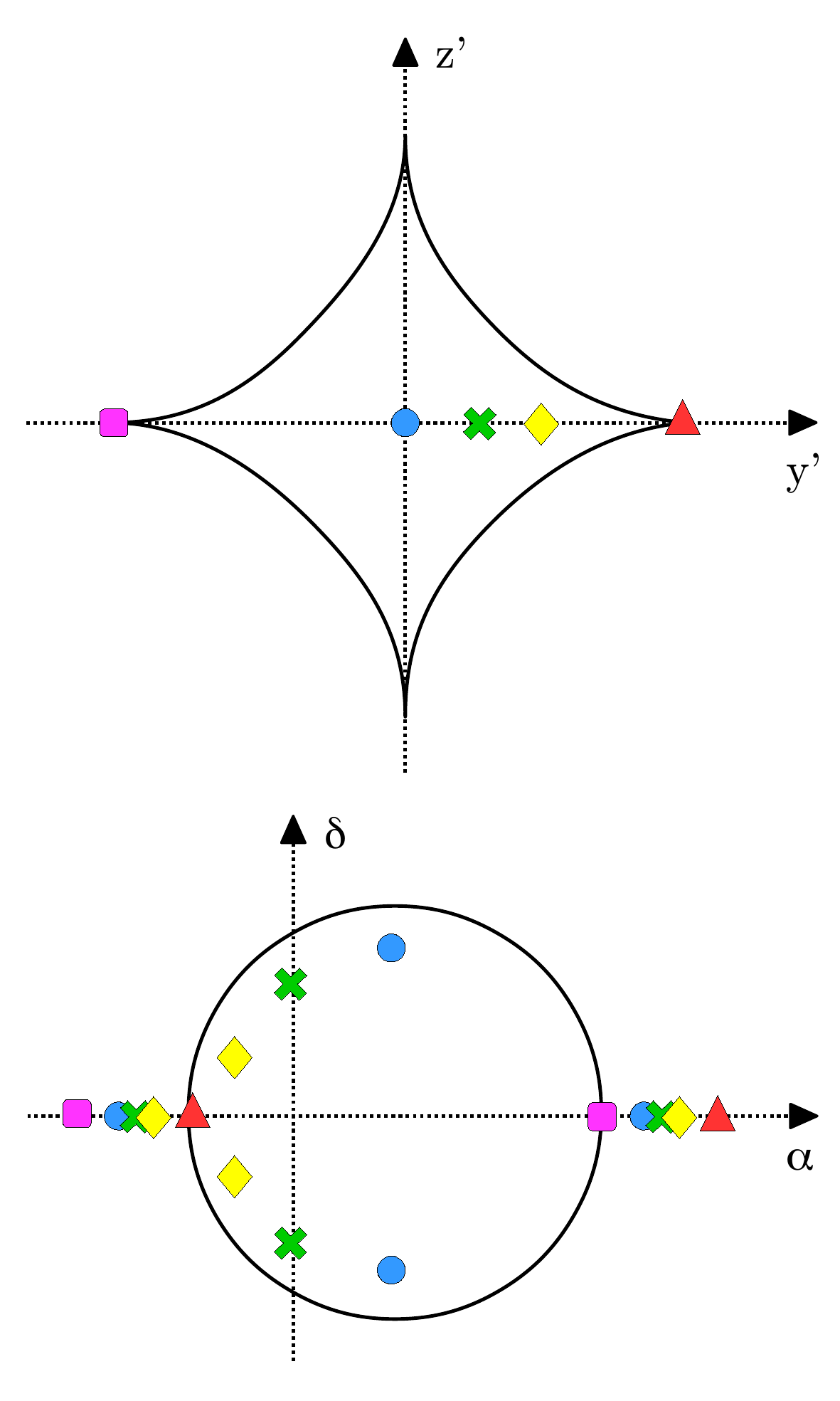}
        \caption{Effect on the displacement of a point source, in the equatorial plane of the black hole, with respect to the caustic (top) on the formation of images in the observer plane (bottom). The black astroid represents the primary caustic and the black circle represents the primary critical curve. We define the $z'$ and $y'$ axis whose origin corresponds to the middle of the astroid cross-section. The $\alpha$ and $\delta$ axis correspond to the observer's sky axis, the origin is centered in the middle of the screen (because of the rotation of the black hole, the critical curve is shifted from the center of the screen).}
        \label{fig:C_CC}
\end{center}
\end{figure}

In case of the Kerr black hole , there is also a primary caustic and higher-order caustics but these are no longer lines  \citep{1994ApJ...421...46R, 2008PhRvD..78f3014B}. \cite{1994ApJ...421...46R} were the first to discover that the primary caustic is a tube with an astroid (four-cusped) cross-section (see the upper scheme in Fig. \ref{fig:C_CC}). At large distances the cross-section is symmetric but becomes distorted near the horizon. In addition, the closer to the BH the source is, the larger the tube shifts with respect to the Schwarzschild's case.
For sources near the horizon, the caustic winds around the black hole in the opposite sense with respect to its rotation. Very far from the black hole, the shift is still significant but the size of the caustic (distance between the right and the left cusp of the astroid cross-section) decreases and tends towards zero.

To form the critical curve the source must cover all of the astroid cross-sections. If we consider a point source in the caustic, four images are formed in the observer's sky. If the point source is on the caustic surface or outside the caustic, only two images are formed.  An illustration is given in Fig. \ref{fig:C_CC}.
Because of the rotation of the black hole, if the source lies on the right cusp one of the two images will be formed on the left side of the critical curve.

\begin{figure*}[!t] 
\centering
      \includegraphics[scale=0.4]{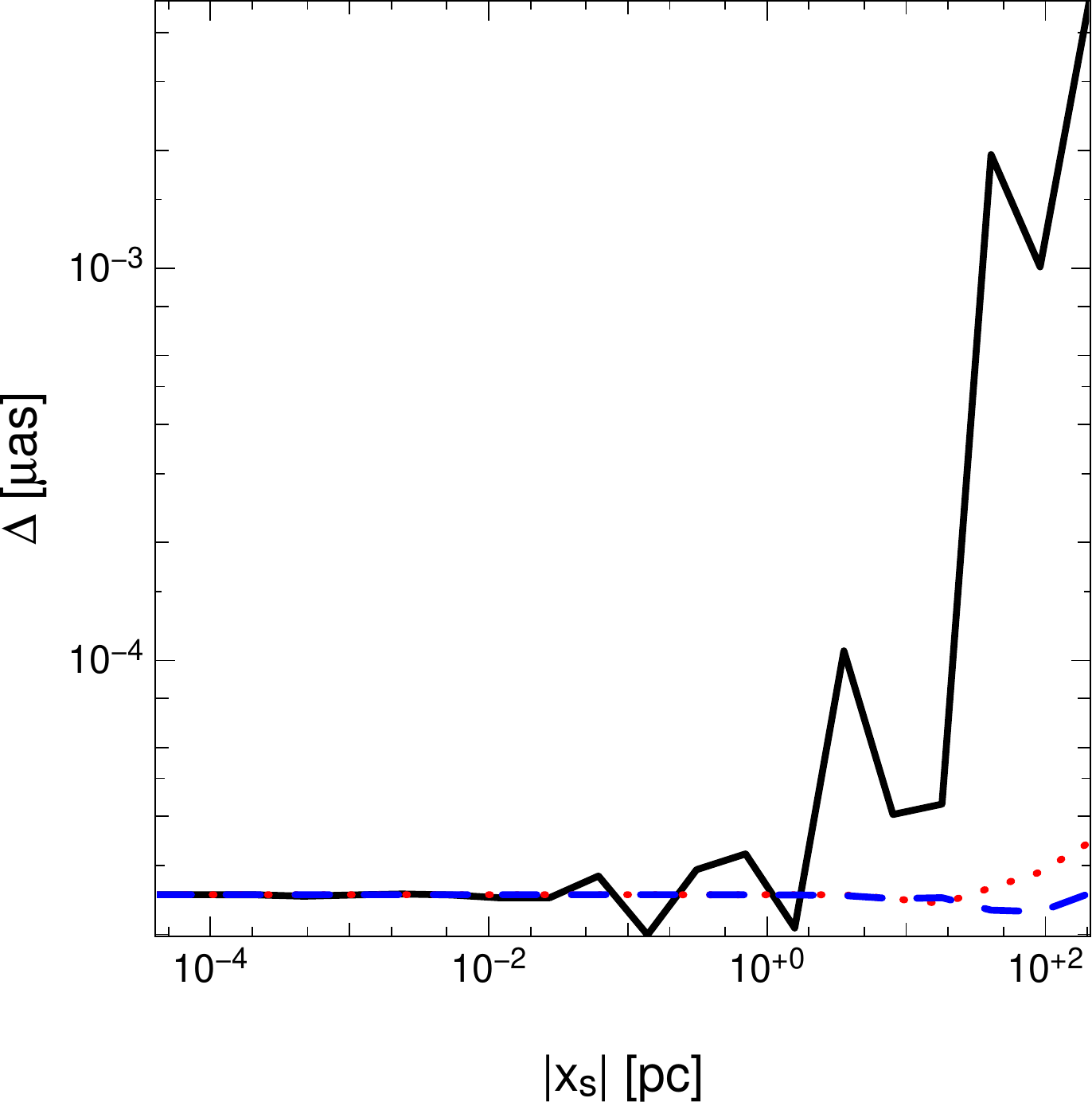}
      \hfill
      \includegraphics[scale=0.4]{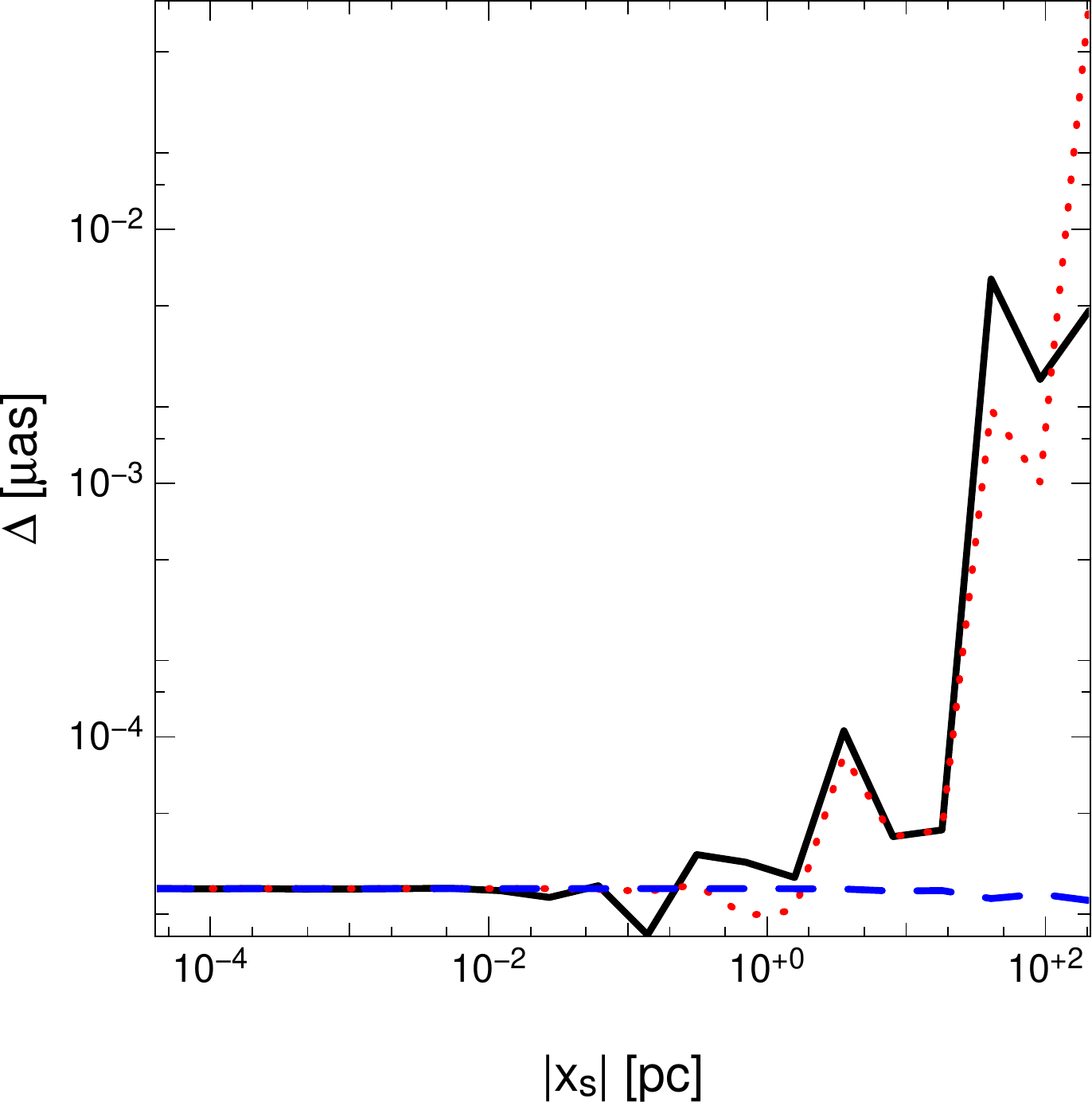}
      \hfill
      \includegraphics[scale=0.4]{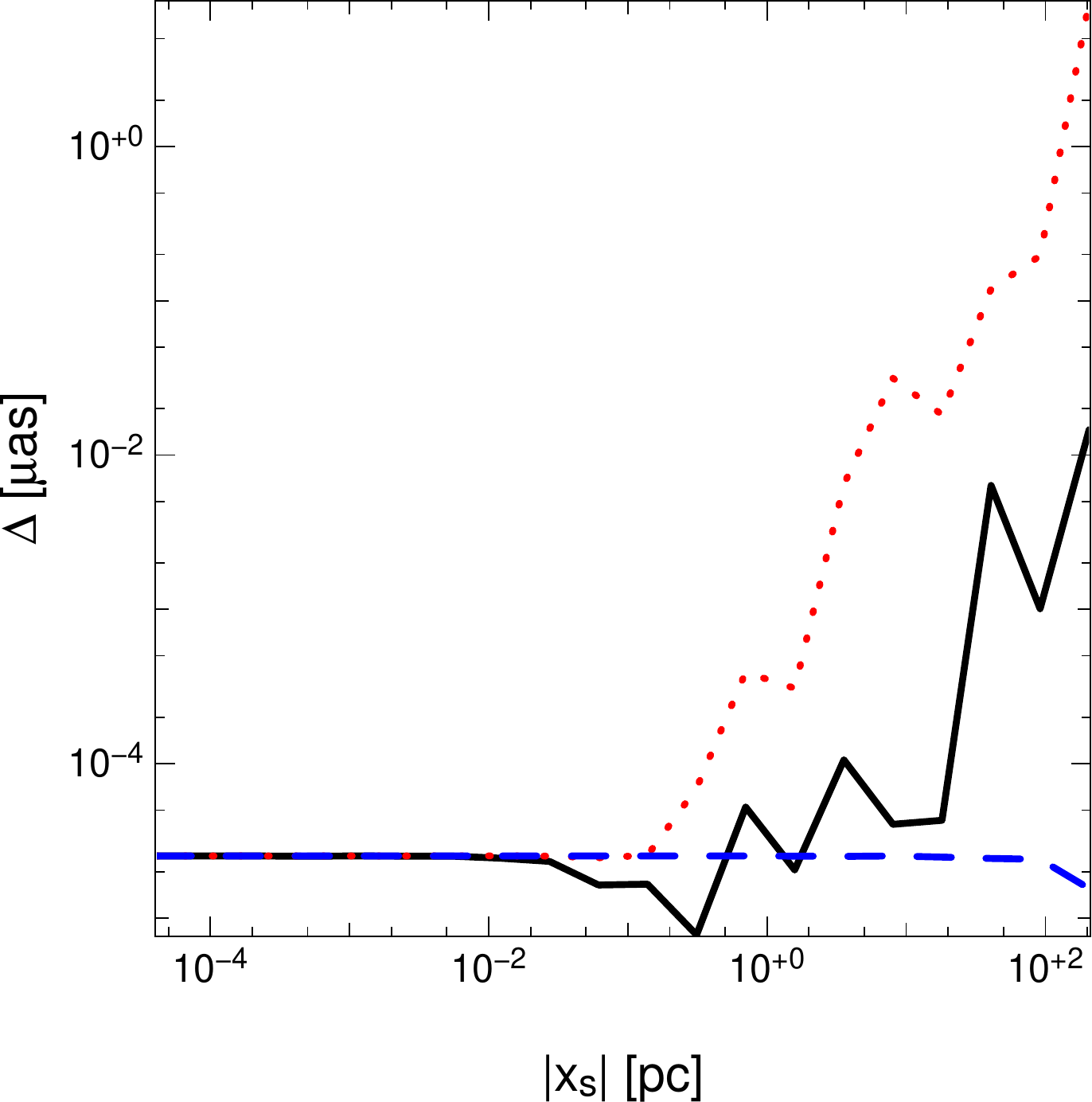}
   \caption{Offset estimated with GYOTO in $\mu$as versus the distance $|x_s|$ of the point source in parsec, in the Schwarzschild case. Each plot represents different values of the tuning parameters $h$ and \texttt{AbsTol}: $10^{-18}$ (left),  $10^{-16}$ (middle), and $10^{-14}$ (right). The different types of curves denote the different integrators:  \texttt{Legacy Generic Integrator} (solid), \texttt{Runge Kutta Cash Karp 54} (dash), and \texttt{Runge Kutta Fehlberg 78} (dot).}
        \label{fig:AT}
\end{figure*}

Analytical studies of primary caustics and primary critical curves has been made  \citep{2006PhRvD..74l3009S, 2008PhRvD..78b3008S}. All of their analytical approximations were obtained in both weak-deflection and weak-field regimes. In the WDL, photons do not wind around the black hole which means that the minimum distance between the photon and the lens $r_{\mathrm{min}_{\gamma}}$ must satisfy $R_S \ll r_{\mathrm{min}_{\gamma}}$. In the weak-field regime, we verify the condition $r_s \gg R_S$. In both regimes, the primary caustic is only shifted and keeps a symmetric shape. Because of the shift of the caustic, the critical curve is not centered on the black hole. In \cite{2008PhRvD..78b3008S}, three equations are developed through a Taylor expansion of the null geodesics in
\begin{equation}
        \varepsilon = \frac{\theta_E}{4D},
\end{equation} 
where $D = r_s / (r_s+r_0)$ and $\theta_E$ is the Einstein ring radius. 
All of the equations  in this paper are expressed in the equatorial plane of the black hole. One of the three equations gives the radius of the critical curve:
\begin{align}
 \Theta_E &\approx \theta_E \Bigg\{ 1 + \frac{15\pi}{32}\varepsilon +   \left[4\left(1+D^{2}\right)  - \frac{675\pi^{2}}{2048} \right] \varepsilon^{2} \notag \\ 
& + \frac{15\pi}{8}\varepsilon^{3}  \left[  D + 4D^{2} -   \frac{9(272-25\pi^{2})}{1024} -  \frac{a^{2}}{8} \right]  \Bigg\}.
\label{eq:theta}
\end{align}
The left and the right radius of the critical curve are equal and depend on the spin in the third-order term in $\varepsilon$.
The second equation gives the offset $\Delta$ of the center of the critical curve relative to the black hole:\begin{equation}
        \Delta \approx 4D\varepsilon^{2} a \Bigg\{ 1 + \frac{15\pi}{32}\varepsilon + \left[ 4(2 - D + 4D^2) - \frac{225\pi^2}{256} \right] \varepsilon^{2} \Bigg\}.
        \label{eq:delta}
\end{equation}
Because the black hole is spinning from the left to the right, the critical curve is shifted to the right (y>0).
The last equation gives the position $y_C$ of the primary caustic for a given $x_s$ and $z_s:$
\begin{align}
        y_C \approx \frac{a}{1-D} \left[  1+ \frac{5\pi}{16} \varepsilon + \left(4 - \frac{225\pi^2}{512}\varepsilon^2 \right) \right]  + \Delta_C\cos{^3\eta},
\end{align}
with 
\begin{align}
        \Delta_C \approx \frac{15\pi}{256} \frac{a^2}{D(1-D)r_0}, 
\end{align}
where $\Delta_C$ is the size of the caustic. The angle $\eta$ ranges from 0 to $2\pi$. In this study, we decide to focus on the right cusp of the primary caustic and to determine the angular position $B_C$ of this cusp seen from the Earth. It means that we assume $\eta = \pm \pi$ and $z_s = 0$. This yields, for a given $x_s$
\begin{equation}
B_C \approx \arctan{\frac{y_C}{x_0+|x_s|}}.
\label{eq:ang}
\end{equation}

Analytical approximations in the SDL have also been derived \citep{2005PhRvD..72h3003B,2007PhRvD..76h3008B} but not for the primary caustic since it is formed in the WDL. A numerical study was made by \cite{2008PhRvD..78f3014B} who study the full structure of caustics and critical curves and compared the results with available analytical formulas. In the case of primary caustics, the author compared the size and the position of the left cusp of the caustic and showed that the approximations of Sereno and De Luca only fail at very small distances from the horizon. \\

To validate GYOTO in the WDL, we estimate the three parameters $\Theta_E$, $\Delta,$ and $B_C$ with GYOTO and make a comparison with formulae $\eqref{eq:theta}$, $\eqref{eq:delta}$, and $\eqref{eq:ang}$, respectively. To reproduce the observational conditions of GRAVITY, we consider an observer at  $r_0 = 8$ kpc from a black hole of mass $M$ equal to $4.31 \times 10^{6} M_{\odot}$. 
We also consider a source far enough from the black hole to be compliant with the domain of validity of these approximations.
In the next two sections, we estimate the three parameters as a function of the position $x_s$ of the source. The methods used to measure these parameters using GYOTO are given in Appendix~\ref{app:A}. 
For each distance of the source, we estimate the error made on the different parameters. 
Since the goal of this paper is to determine whether the accuracy of GYOTO is better than 1~$\mu$as, we only consider the maximum error for each parameter (see for instance the end of Appendix~\ref{app:A11} for the maximal error estimation of the angular position of the caustic point).
The next section is devoted to  discussing  the different integrators implemented in GYOTO and used to integrate null geodesics.


\section{Comparison of various integrators of GYOTO}
\label{sec:three}

\begin{figure*}[!t]
\centering 
      \includegraphics[scale=0.4]{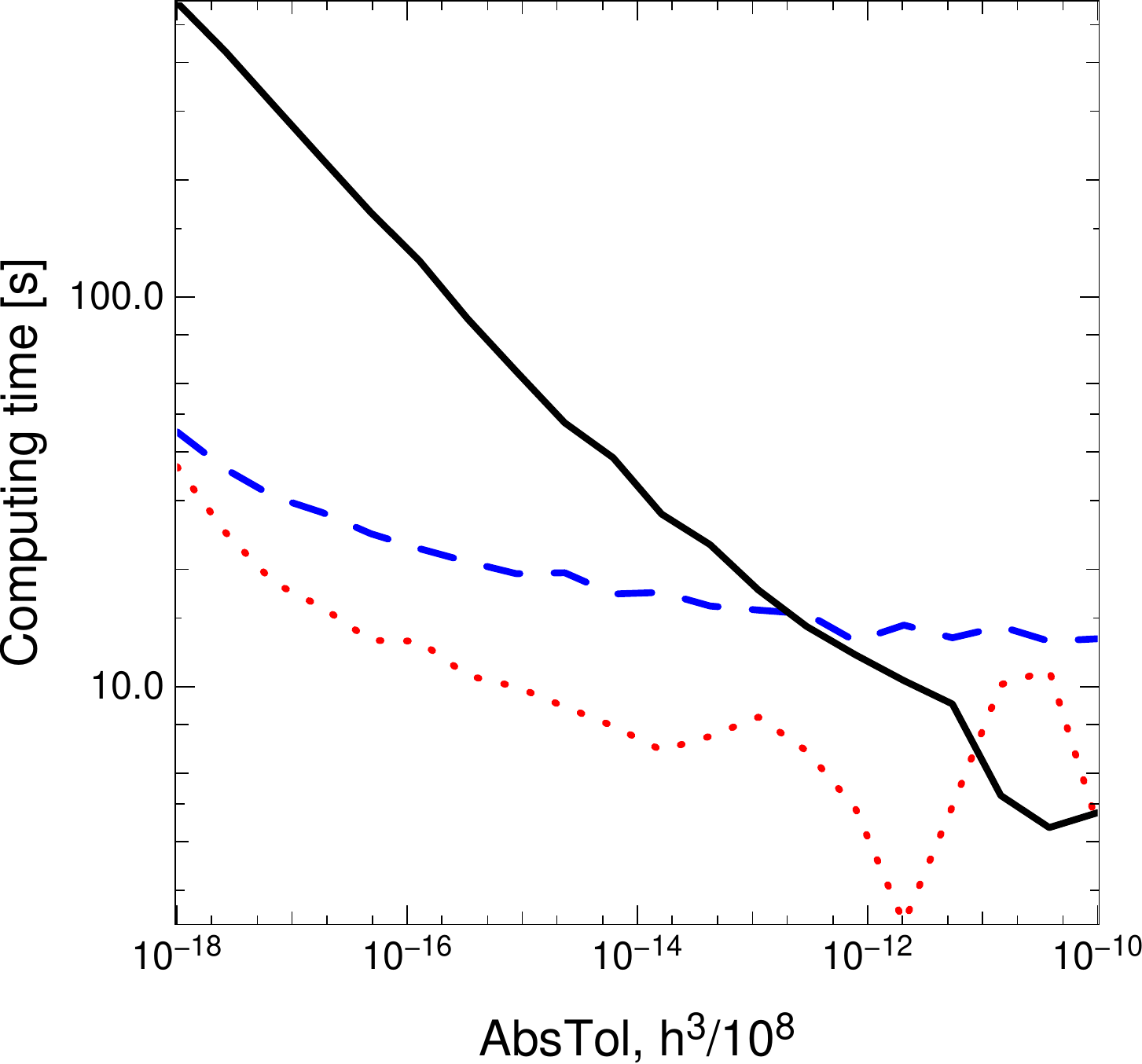} 
      \hfill
      \includegraphics[scale=0.4]{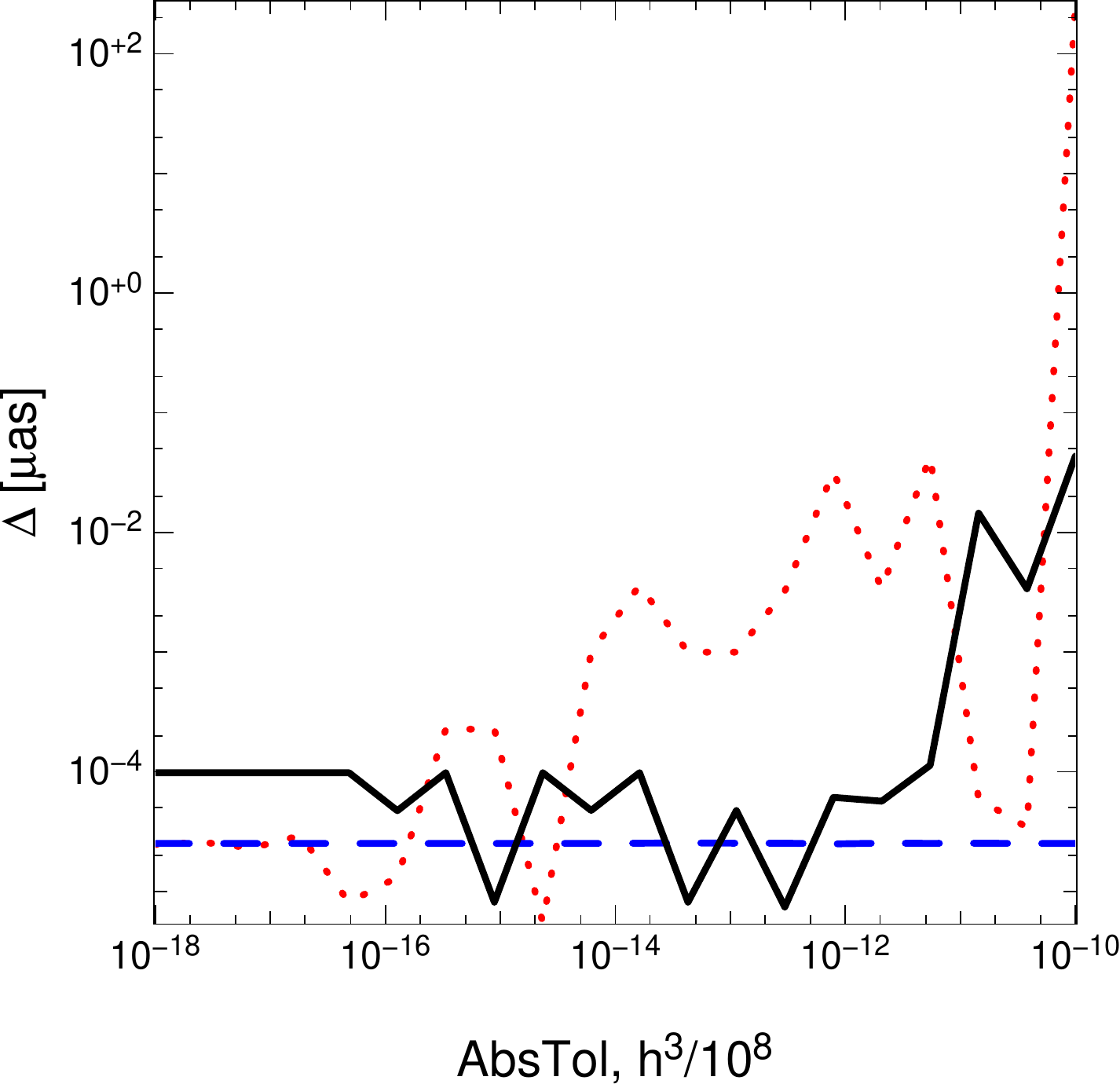}
      \hfill
      \includegraphics[scale=0.4]{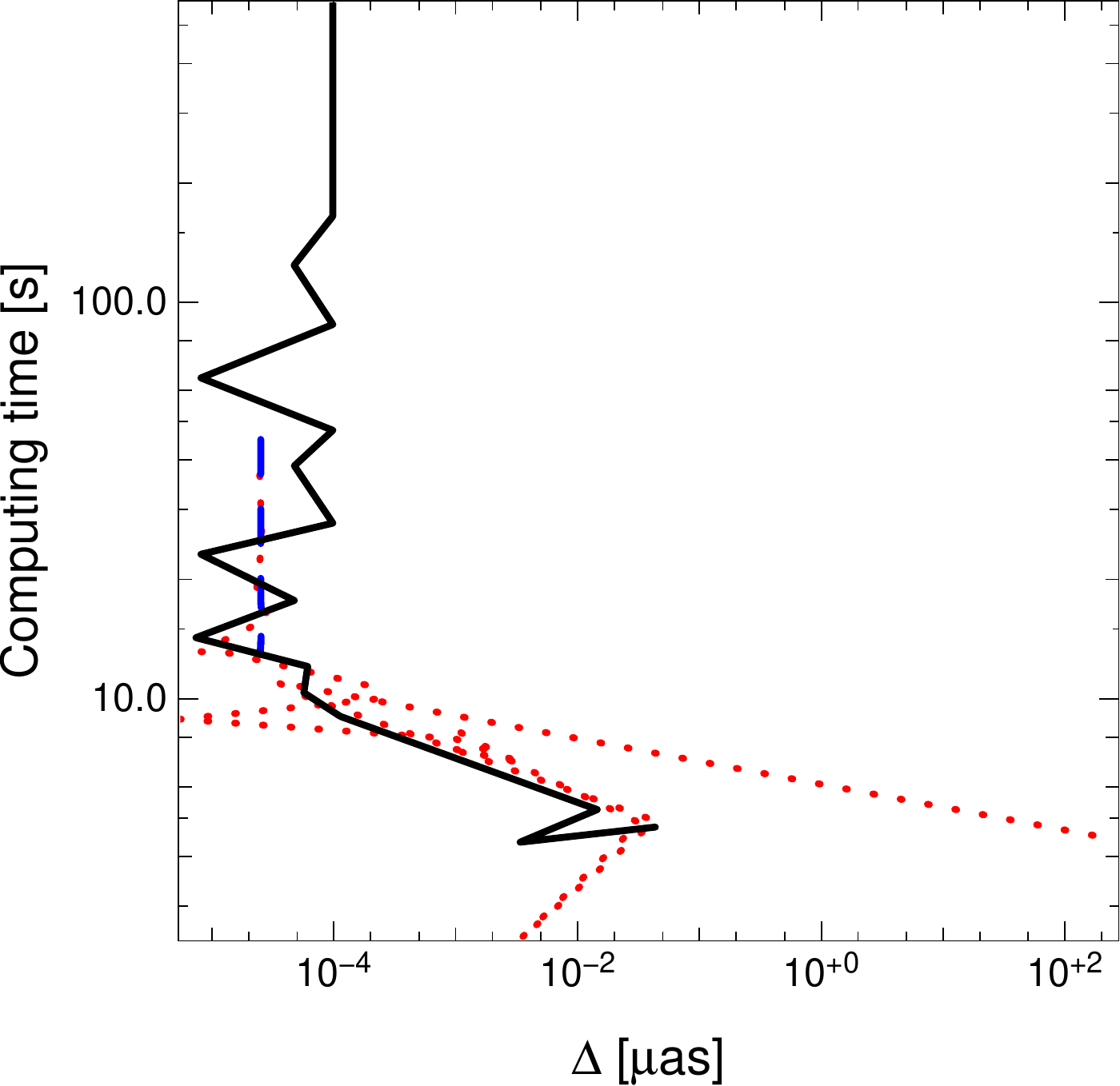}
   \caption{Comparison of the three integrators by measuring the offset of the Einstein ring in the Schwarzschild case. Left: computing time versus the tuning parameters. Middle: offset versus the tuning parameters. Right: computing time versus the offset. The different types of curves denote the different integrators:  \texttt{Legacy Generic Integrator} (solid), \texttt{Runge Kutta Cash Karp 54} (dash), and \texttt{Runge Kutta Fehlberg 78} (dot).}
   \label{fig:TAD}
\end{figure*}

To compute null geodesics in GYOTO, different integrators were implemented to solve the equations of motion of the photon. Before estimating the three parameters, we need to do an intrinsic test of GYOTO to know which integrators are best. To test their validity, we estimate the offset of the critical curve in the Schwarzschild case, which means that $\Delta$ should tend to zero. In this section, we do not estimate the position of the caustic point, we directly put the source at ($x_s$,0,0). For more details on the estimation of the offset see Appendix~\ref{app:A2}. Six different integrators are available in GYOTO, however we choose to compare the following three:
\begin{itemize}
\item the \texttt{Legacy Generic} integrator. This is the first integrator that has been implemented in GYOTO. It is a fourth-order Runge Kutta integration controlled by several parameters. An important tuning parameter for this integrator, \texttt{DeltaMaxOverR} ($h$), depends on the maximal integration step ($ \delta_{\mathrm{max}}$) and the current distance to the center of the coordinate system ($r$) as $h = \delta_{\mathrm{max}} / r$. Thus, the integration step cannot be larger than $h \times r$.
\item the \texttt{Runge Kutta Cash Karp 54} integrator. This is an integrator of the Odeint library (for numerically solving ordinary differential equations) belonging to the Boost C++ Libraries\footnote{see \url{http://www.Boost.org/}}. Several Boost integrators have recently been  added to GYOTO because they are well debugged thanks to their wide user base and enable the user to make run-time choices depending on the problem at hand. Several tuning parameters are used to optimize the integration. For instance, the integrator controls the error on the step integration by computing two steps with different orders, the fourth and the fifth. It estimates the difference between these two steps to evaluate the error. This error is compared against  an error tolerance depending on two parameters. They are called \texttt{AbsTol} and \texttt{RelTol} and represent the absolute and the relative tolerance, respectively. 
\texttt{RelTol} is used for high values of solutions and \texttt{AbsTol} for solutions close to zero.
\item the \texttt{Runge Kutta Fehlberg 78} integrator. This is another integrator from the Boost family. It computes the two steps with higher orders, the seventh and the eighth.
\end{itemize}
The offset is obtained taking into consideration these three integrators and different values for the tuning parameters $h$, \texttt{AbsTol,} and \texttt{RelTol}. We choose to work with $\texttt{RelTol} = \texttt{AbsTol}$ but this is not a unique choice. We did not investigate independent variations of the two parameters because this is not relevant to our study. In what follows, the two parameters are set equal and we only use the name \texttt{AbsTol}.

We also note that there is a systematic error in GYOTO generated by the machine accuracy. More precisely, this bias corresponds to the position of the observer, which is not exactly at $y_s=0$ but is shifted by approximately $y_s=10^{-6} M$ owing to the choice of coordinate system. This shift leads to a (negligible) systematic effect around  $10^{-5} \mu$as in the observer plane.

The offset of the critical curve for $h^3/10^8$ and \texttt{AbsTol} equal to $10^{-18}$, $10^{-16}$, and $10^{-14}$, versus the absolute distance of the source $|x_s|$ in parsec (the source is always behind the black hole thus $x_s$<0) is presented in Fig. \ref{fig:AT}. We consider  varying $h^3/10^8$ and not $h$ for the \texttt{Legacy} integrator because the tuning parameters \texttt{AbsTol} and $h$ are not defined in the same way. The quantity $h^3/10^8$ is obtained empirically to globally recover the same error behavior between \texttt{Legacy} and Boost integrators.
In Fig. \ref{fig:AT}, the noise generated by GYOTO increases with the distance of the source to the black hole for all integrators. Boost integrators are more accurate than \texttt{Legacy} for tuning parameters equal to $10^{-18}$. However, the offset of the \texttt{Legacy} integrator remains small ($< 10^{-2} \mu$as), even for large distances and for the three values of $h$. The offset obtained with the \texttt{Runge Kutta Fehlberg 78} integrator significantly increases with the distance for \texttt{AbsTol} =  $10^{-16}$ and $10^{-14}$, and can exceeds 1 $\mu$as for very large $|x_s|$. The other Boost integrator \texttt{Runge Kutta Cash Karp 54} is accurate ($\Delta \approx 10^{-5} \mu$as) for all the distances investigated and for the three values of \texttt{AbsTol}. 

When the star is located between $|x_s| \approx 10^{-4}$ parsec and $|x_s| \approx 10^{-1}$~parsec, the offsets $\Delta$, which are evaluated for each integrator and each value of tuning parameter, fluctuate around $10^{-5}$$\mu$as. These errors are dominated by the observer position shift that is due to machine accuracy. However, a second error appears for larger distances of the star and is due to the interpolation made in GYOTO. More precisely, this interpolation is made in a function called \texttt{MinDistance} in GYOTO. This function is used to evaluate the right and left Einstein radii needed to estimate the offset $\Delta$ (see Appendix~\ref{app:A} for the definition of the \texttt{MinDistance} function and an explanation on the method implemented to obtain $\Delta$). This limitation is due to the step size in which the interpolation is made to estimate the \texttt{MinDistance} function. Indeed, when the step size increases, the interpolation is less efficient and the offset $\Delta$ increases. This second limitation is only apparent for the \texttt{Legacy} and \texttt{Runge Kutta Fehlberg78} integrators because these two integrators generate less intermediate points in this context, and therefore incur more interpolation. This source of error is therefore not cumulative (it is reset at each point estimated by the integrator). We have looked at the statistics for this error, which are essentially Gaussian.\\

\begin{figure*}[!t]
\centering 
      \includegraphics[scale=0.4]{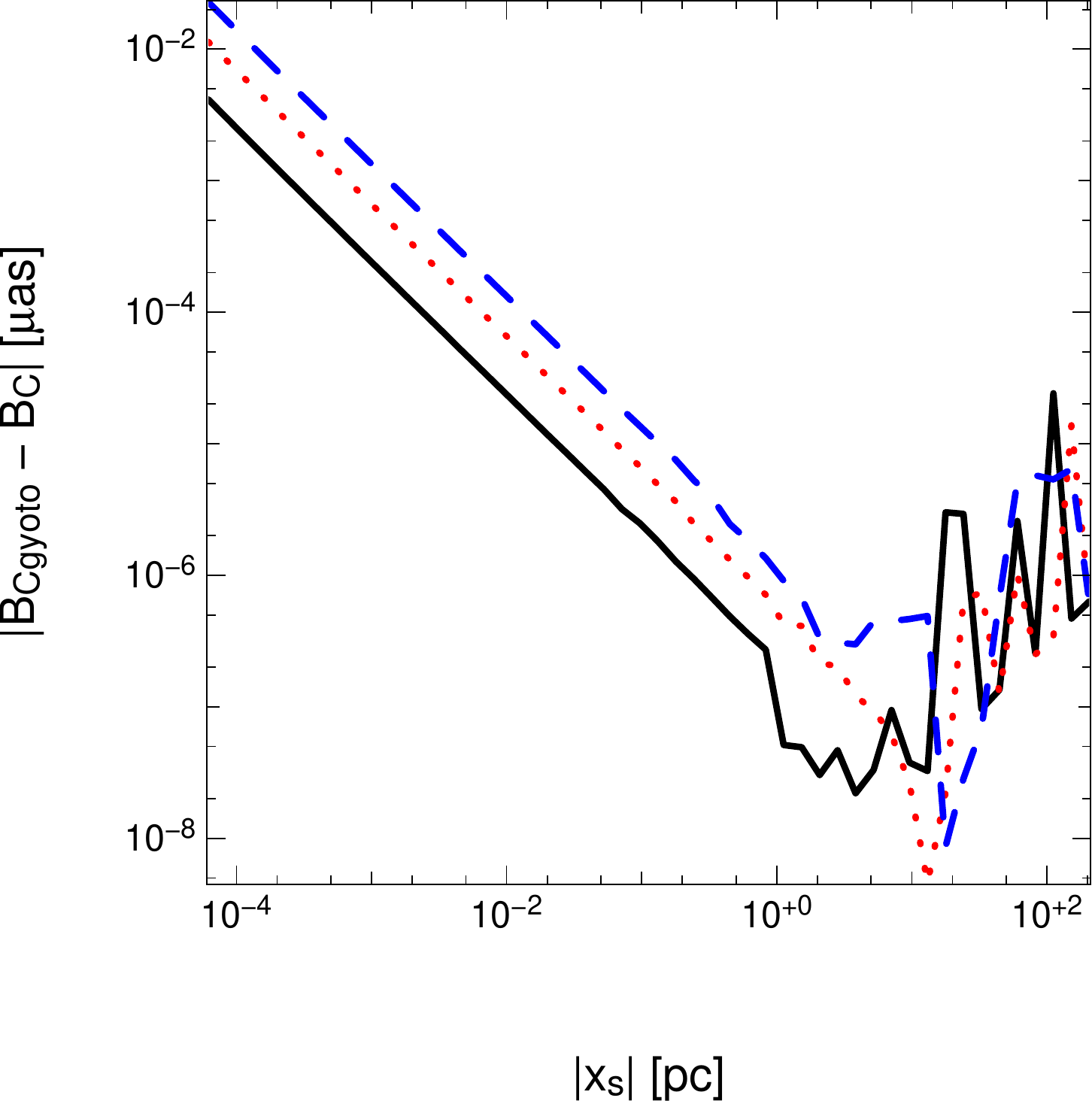}
      \hfill
      \includegraphics[scale=0.4]{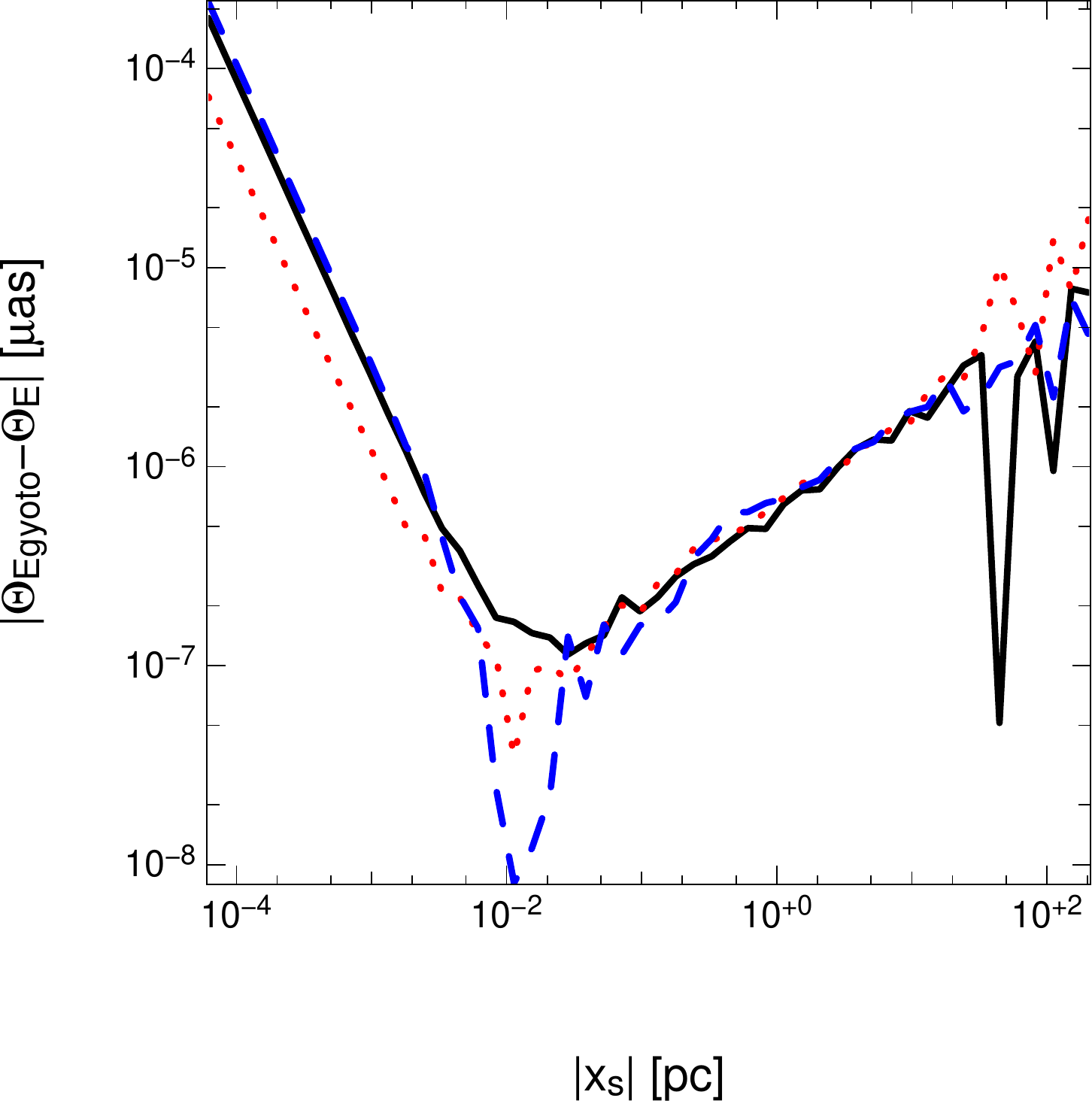}
      \hfill
      \includegraphics[scale=0.4]{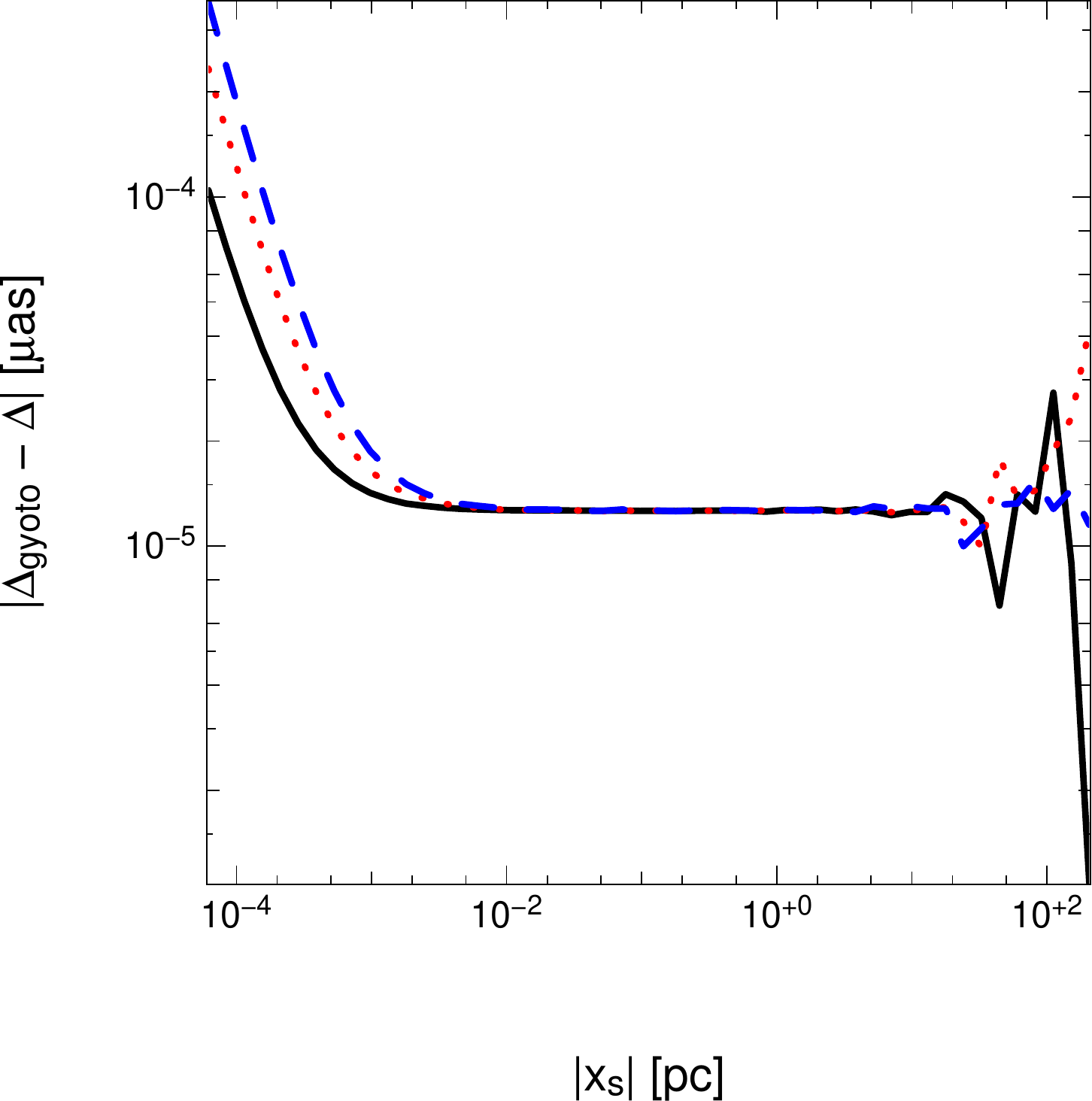}
   \caption{Absolute difference between analytical approximations  $\eqref{eq:theta}$, $\eqref{eq:delta}$, and $\eqref{eq:ang}$, and GYOTO measurements for the three parameters $B_C$, $\Theta_E,$ and $\Delta$. The types of line denote different values of the spin: 0.2 in solid, 0.5 in dotted, and 0.9 in dashed.}
   \label{fig:ATD}
\end{figure*}

All integrators seem to meet the requirements for the studies to follow. 
However, we also decided to evaluate the computing time needed by the various integrators to reach their best astrometric accuracy. To do so, we estimate the offset of the Einstein ring for different values of the tuning parameters (ranging from $10^{-10}$ to $10^{-18}$), and the computing time needed by each integrator to evaluate the offsets. We used a 2.66 GHz Intel Core 2 Duo processor to compute null geodesics. We also consider a star at $|x_s| = 2$ parsecs.

Figure \ref{fig:TAD} shows that \texttt{Runge Kutta Fehlberg 78} is the fastest integrator for all values of the tuning parameters. Its best accuracy ($\approx 10^{-5} \mu$as) is reached for \texttt{AbsTol} in the range $10^{-17}$--$10^{-18}$, corresponding to a computing time in the range $15$--$35$~s. The \texttt{Runge Kutta Cash Karp 54} integrator is always accurate ($\approx 10^{-5} \mu$as), its computing time is between 15 seconds (\texttt{AbsTol}  $= 10^{-10}$) and 45 seconds (\texttt{AbsTol}  $= 10^{-18}$). The best accuracy for the last integrator ($\approx 10^{-5} - 10^{-4} \mu$as) is reached for $h^3/10^8$ between $10^{-18}$ to $10^{-12}$, which means a computing time between 10 minutes to 10 seconds, respectively. The optimal compromise between precision and computing time is thus reached at $\texttt{AbsTol} = 10^{-17}$ for the \texttt{Runge Kutta Fehlberg~78} integrator, $\texttt{AbsTol} = 10^{-10}$ for \texttt{Runge Kutta Cash Karp54} and $h^3/10^8 = 10^{-12}$ for \texttt{Legacy}, where the offset is around $10^{-5} \mu$as and the time needed is less than one minute.

The important time difference between Boost integrators and the \texttt{Legacy} integrator at $\texttt{AbsTol} = 10^{-18}$ is due to the method used to estimate the adaptive step during the integration. Indeed, considering the number of steps used to estimate one null geodesic, $\approx$ 90 000 steps are needed for the \texttt{Legacy} integrator, however, Boost integrators only need one order of magnitude less than \texttt{Legacy}.
The accuracy limitation on $\Delta$ on this figure is also dominated by the two types of error discussed above.\\

This quick study shows that the three integrators are appropriate for the following studies. In addition to the optimal compromise mentioned above, other reasonable choices of numerical parameters can be made such as $\texttt{AbsTol} = 10^{-18}$ with \texttt{Runge Kutta Fehlberg 78}, where we still have an accuracy of around $10^{-5} \mu$as and a computing-time of less than one minute. Since the following studies do not use time-consuming methods, we choose to consider $\texttt{AbsTol} = 10^{-18}$ and \texttt{Runge Kutta Fehlberg 78}. This is the choice that we made for the rest of this paper.


\section{Results}
\label{sec:four}


\subsection{Validation of GYOTO in WDL}
\label{ssec:WDL}

In Fig. \ref{fig:ATD}, we present the absolute difference between analytical approximations and GYOTO measurements obtained for three values of the spin, 0.2, 0.5, and 0.9. For the range of parameters investigated, the differences are always extremely small ($\lesssim 10^{-2}\mu$as). 
\begin{figure}[!b]
\centering
     \includegraphics[scale=0.5]{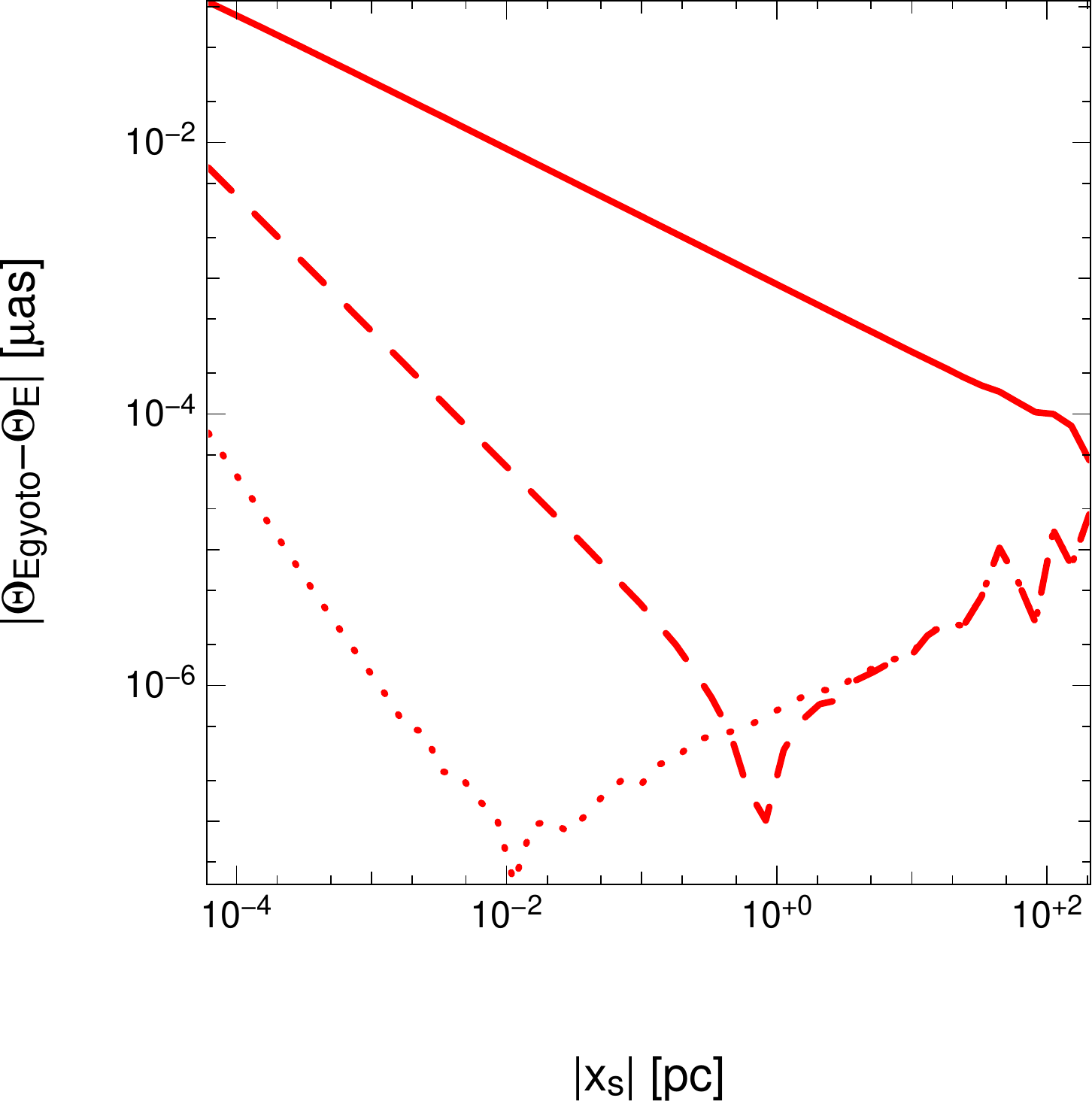}
      \caption{Absolute difference between analytical approximations and GYOTO measurements considering three different equations for the approximation $\Theta_E$: the solid line is obtained considering only the zeroth and first orders in $\varepsilon$ in the equation \eqref{eq:theta}, the dashed line considers the zeroth, first, and second orders in $\varepsilon$ and the dotted line considers the full equation. Only $a$ = 0.5 is plotted.}
   \label{fig:Th_orders05}
\end{figure}
On each plot two different regimes can be observed. For small $|x_s|$, the curve is marked by a smooth, power-law decrease: GYOTO and the numerical approximation agrees better and better for larger and larger values of $|x_s|$. After reaching a minimum, the curve raises again, with a much more noisy appearance. 
This is due to GYOTO being better than 
analytical approximations  for small $|x_s|$. The difference between the two traces the order of the approximation. On the other hand, for large values of $|x_s|$, the approximations win over GYOTO and the difference is dominated by the numerical error of GYOTO. To demonstrate this interpretation, we compare GYOTO to three approximations of different order in Fig.~\ref{fig:Th_orders05}.
The first absolute difference is obtained considering the zero and the first orders in $\varepsilon$ ($\varepsilon^0$ and $\varepsilon$) in Equation \eqref{eq:theta}. For the second absolute difference, we consider one more order in $\varepsilon$ ($\varepsilon^0$, $\varepsilon$ and $\varepsilon^2$). The last one is obtained considering the full Equation \eqref{eq:theta}, which is the same plot shown in the middle of Fig. \ref{fig:ATD}. We can see that, when smaller orders are considered, the differences decrease for all distances of the source. We also note that the noisy part appears earlier and earlier when considering accurate approximations. We thus confirm that the first part of the curves presented in Fig.\ref{fig:ATD} is due to analytical approximations.

The maximal errors on each parameter evaluated with GYOTO, and for each spin, are presented in Table~\ref{table:WDL}. 
The method to estimate this error is presented in Appendix~\ref{app:A2}.
All the maximal errors in Table~\ref{table:WDL} are around $10^{-5}$~$\mu$as. The limitation in the estimation of these errors is due to the systematic bias induced in the observer screen. The ray-tracing code is very accurate, even for sources far behind the black hole (e.g. $\delta_{\Theta_{E\mathrm{gyoto}}} \approx 10^{-5} $~$\mu$as at 200 parsecs). \\

\begin{table}[!h]
\begin{center}
\begin{tabular}{|c|c|c|}
\hline
        $a$ & $\delta_{B_{C\mathrm{gyoto}}}$ & $\delta_{\Theta_{E\mathrm{gyoto}}}$, $\delta_{\Delta_{\mathrm{gyoto}}}$  \\[2pt]     
        \hline 
        0.2   & $5.6 \times 10^{-5}$ & $8.2 \times 10^{-5}$ \\[2pt]     
        0.5   & $1.6 \times 10^{-4}$ & $1.1 \times 10^{-4}$ \\[2pt] 
        0.9 & $2.1 \times 10^{-5}$ & $3.9 \times 10^{-5}$ \\[2pt]
        \hline 
\end{tabular}
\end{center}
        \caption{Maximal errors evaluated for each parameter and each spin in $\mu$as.}
        \label{table:WDL}
\end{table}

The requirement on accuracy ($\leqslant 1 \mu$as) is largely met in the WDL.  
We note that our method (see Appendix~\ref{app:A}) uses photons both inside and outside of the equatorial plane of the black hole. Therefore we have demonstrated our conclusion for any photon in the WDL, not only in the equatorial plane. However, an equivalent test is necessary in the SDL.


\subsection{Comparison of GYOTO null geodesics with Geokerr in SDL}
\label{ssec:SDL}

The aim of this subsection is to check if null geodesics computed with GYOTO in the SDL are accurate enough. To do so, we decided to compare photon trajectories computed with GYOTO  with those computed with the ray-tracing code Geokerr. Contrary to GYOTO, Geokerr computes photon coordinates semi-analytically, reducing the equations of motion expressed in the Hamiltonian formulation to Carlson elliptic integrals. 

\begin{figure}[!h]
\begin{center}
        \includegraphics[scale=0.32]{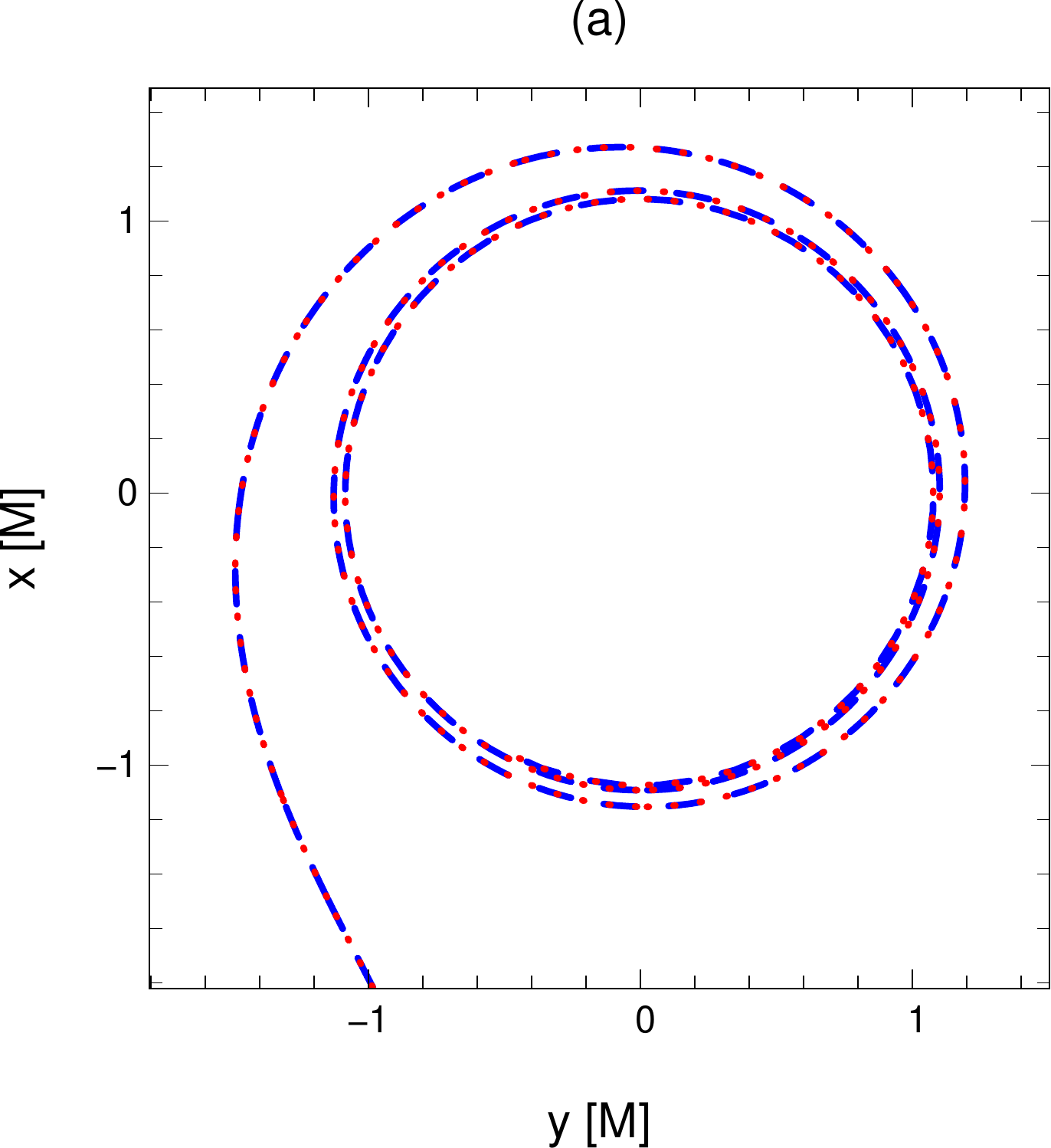}
        \includegraphics[scale=0.32]{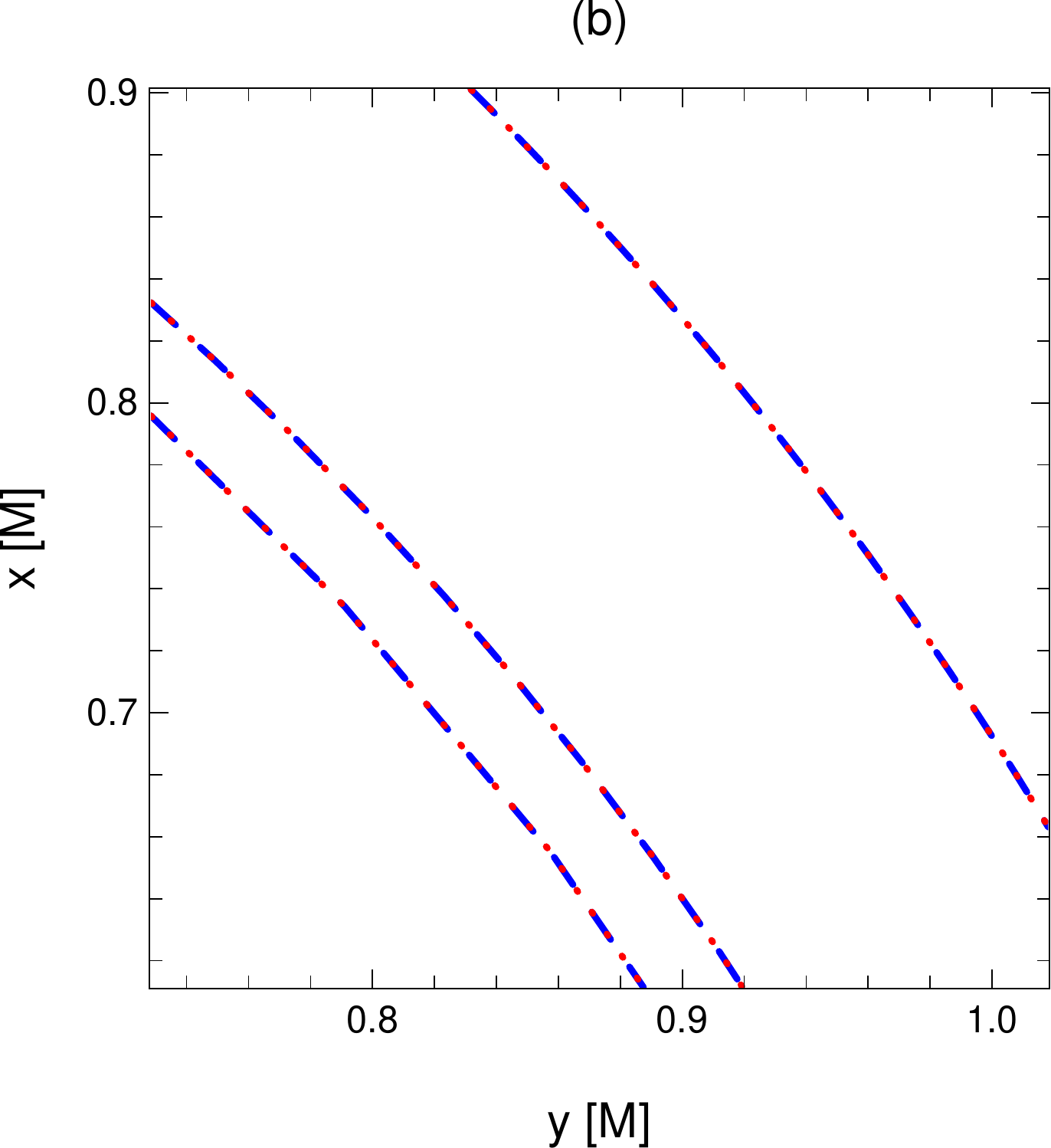}
        \caption{Null geodesics computed with GYOTO (dotted lines) and Geokerr (dashed lines). We consider a spin of 0.998 and a photon launched from the center of the screen. (b) is a zoom of geodesics plotted on (a): $\delta_x$ and $\delta_y$ on (b) is about $4 \times 10^{-4} M$. }
                \label{fig:ph0}
\end{center}
\end{figure}

The comparison is made using the same observer coordinates and black hole parameters as before. We evaluate null geodesics for three different values of the spin (0.2, 0.5 and 0.998) and we consider photons launched from the center of the observer screen ($\alpha=\delta=0$, see the beginning of Sect. 1 of the Appendix~\ref{app:A} for explanation of GYOTO null geodesics integrations). We first compute the geodesics with Geokerr and get the time coordinate of each point of the photon trajectory. Then, to get the null geodesics with GYOTO, we interpolate the positions of photons with our ray-tracing code, taking these dates into consideration. The maximal errors found on the photon coordinates is presented in Table~\ref{table:SDL}. 
\begin{table}[!t]
\begin{center}
\begin{tabular}{|c|c|}
\hline
        $a$ &  $\max(\delta_x,\delta_y,\delta_z)$ \\[2pt]       
        \hline
        0.2   &  $1.7 \times 10^{-3}$ \\[2pt]   
        0.5   &  $2.6 \times 10^{-3}$ \\[2pt] 
        0.998 & $5.2 \times 10^{-3}$ \\[2pt] 
        \hline
\end{tabular}
\end{center}
        \caption{Maximal errors on the photon coordinates evaluated for each spin in $\mu$as.}
        \label{table:SDL}
\end{table}
Errors correspond to the difference between positions evaluated with GYOTO and those evaluated with Geokerr.
The maximal errors are about $10^{-3}~\mu$as which shows a very good consistency between the two ray-tracing codes. Even for a photon, which is not launched from the center of the screen ($\alpha = 5$~$\mu$as and $\delta = -5$~$\mu$as), with $a=0.998$, we find a very small error ($\approx 4  \times 10^{-3}$~$\mu$as).
An example of null geodesics computed with both codes is shown in Fig. \ref{fig:ph0}.

As mentioned, we use the null geodesic integrated in our ray-tracing code to estimate the geodesic coordinates of GYOTO at Geokerr time coordinates. We noted that when the step size is high, the interpolation is less well evaluated. Thus, considering one fixed position on the null geodesic, each Cartesian coordinate is affected by a random error due to the interpolation in GYOTO. If we now focus on the maximal error along each Cartesian coordinate, we found that this error is constant when varying the tolerance parameter. It thus confirms that the error is only linked to the interpolation made in GYOTO.\\

These results show that GYOTO is accurate to a high level, even for large distances and in both weak- and strong-deflection regimes.
Next, we extend the study of \cite{2012ApJ...753...56B} on lensing effects in the central parsec of our Galaxy and discuss the impact of neglecting lensing effects in future stellar-orbit models on the fitted orbital parameters and the detection of the other effects affecting the orbit.

\section{Short studies on lensing effects in the central parsec}
\label{sec:five}

\begin{figure}[!t]
\centering
     \includegraphics[scale=0.5]{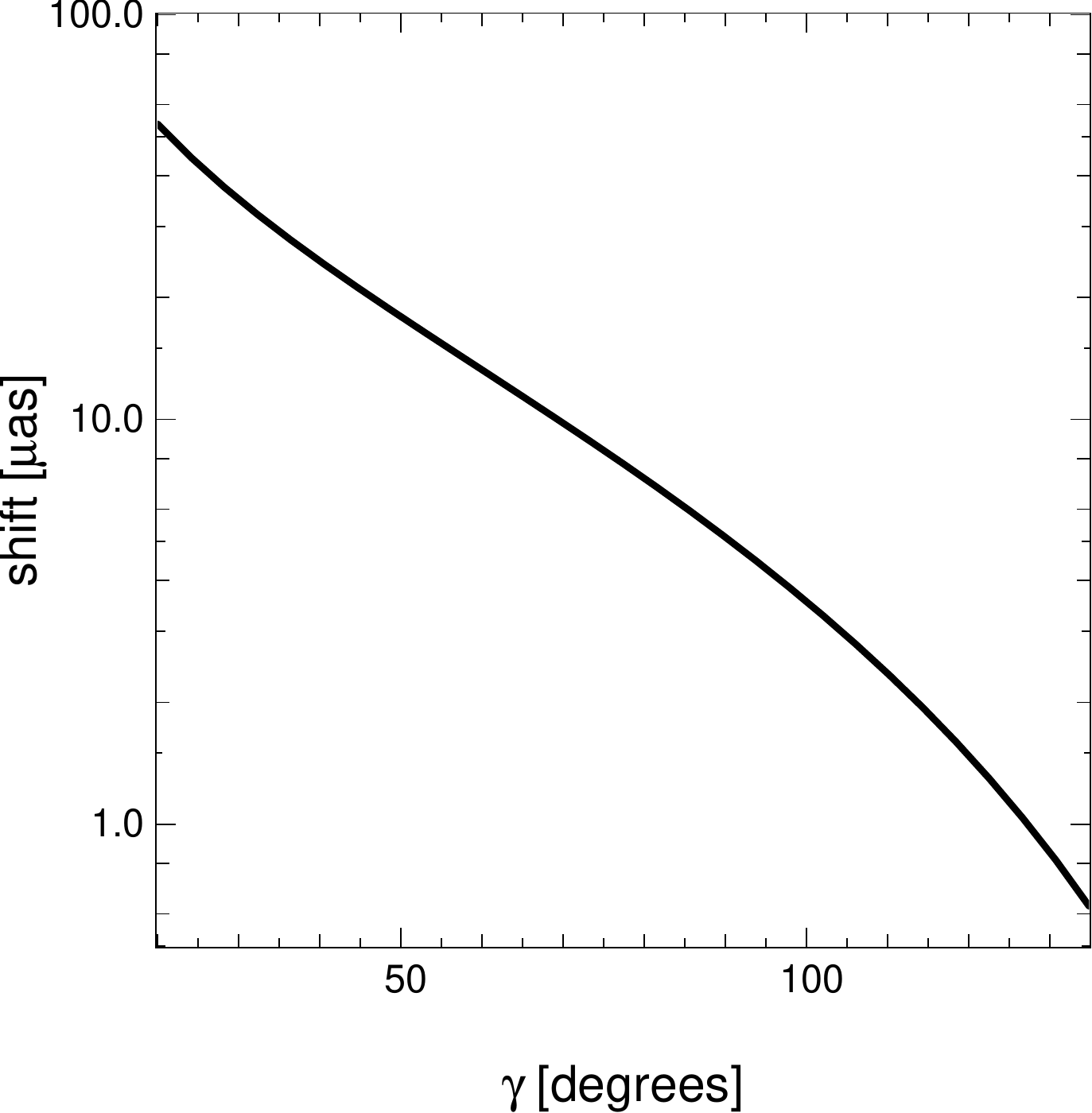} 
      \caption{Astrometric shift in $\mu$as of the primary image owing to lensing effects. This plot is valid for sources ranging from $10^{-4}$ to 1 parsec. The shift is the difference between lensed and unlensed angular positions of the star.}
   \label{fig:shift}
\end{figure}

In this section, we want to complete the study performed by \cite{2012ApJ...753...56B} using the ray-tracing code GYOTO. Considering the astrometric accuracy of GRAVITY, they showed that this instrument is sensitive to the lensing effects generated by a black hole of $4.3 \times 10^{6}~M_{\odot}$. They divided the parameter space in three regions depending on the position of the star. However, they only focused on stars located behind the plane of the sky containing Sgr A*. We present the astrometric shift of the primary image in Fig.~\ref{fig:shift}. We consider stars with $r_s$ ranging from $10^{-4}$ to 1 parsec and vary the angle $\gamma$ corresponding to the angle between the observer, the lens and the source, from $20^{\circ}$ (the star is behind the plane of the black hole) to $135^{\circ}$ (the star is in front of the plane of the sky containing Sgr A*). The shift presented here is obtained in the weak-deflection and weak-field regimes since we satisfy the conditions $r_s \gg R_S$ and $r_{\mathrm{min}_{\gamma}} \gg R_S$. Indeed, the smallest $r_s$ and $r_{\mathrm{min}_{\gamma}}$ we get are of about 250 $R_S$ and 90 $R_S$, respectively.
We consider the same mass of the black hole and distance of the observer ($r_0=8.3 $kpc) as \cite{2012ApJ...753...56B}.

We find the same results as \cite{2012ApJ...753...56B}. Although the effect appears to be small for a star in front of the plane of Sgr A* (e.g. $\approx 2$~$\mu$as at $110^\circ$ inclination), it is a systematic effect which always displaces the image of the star away from the black hole. For this reason, any attempt at orbit-fitting must take the lensing effects into account. For instance, the major axis of an orbit in the plane of the sky will appear $5~\mu$as larger than it really is.\\

\begin{figure}[!b]
\centering
     \includegraphics[scale=0.5]{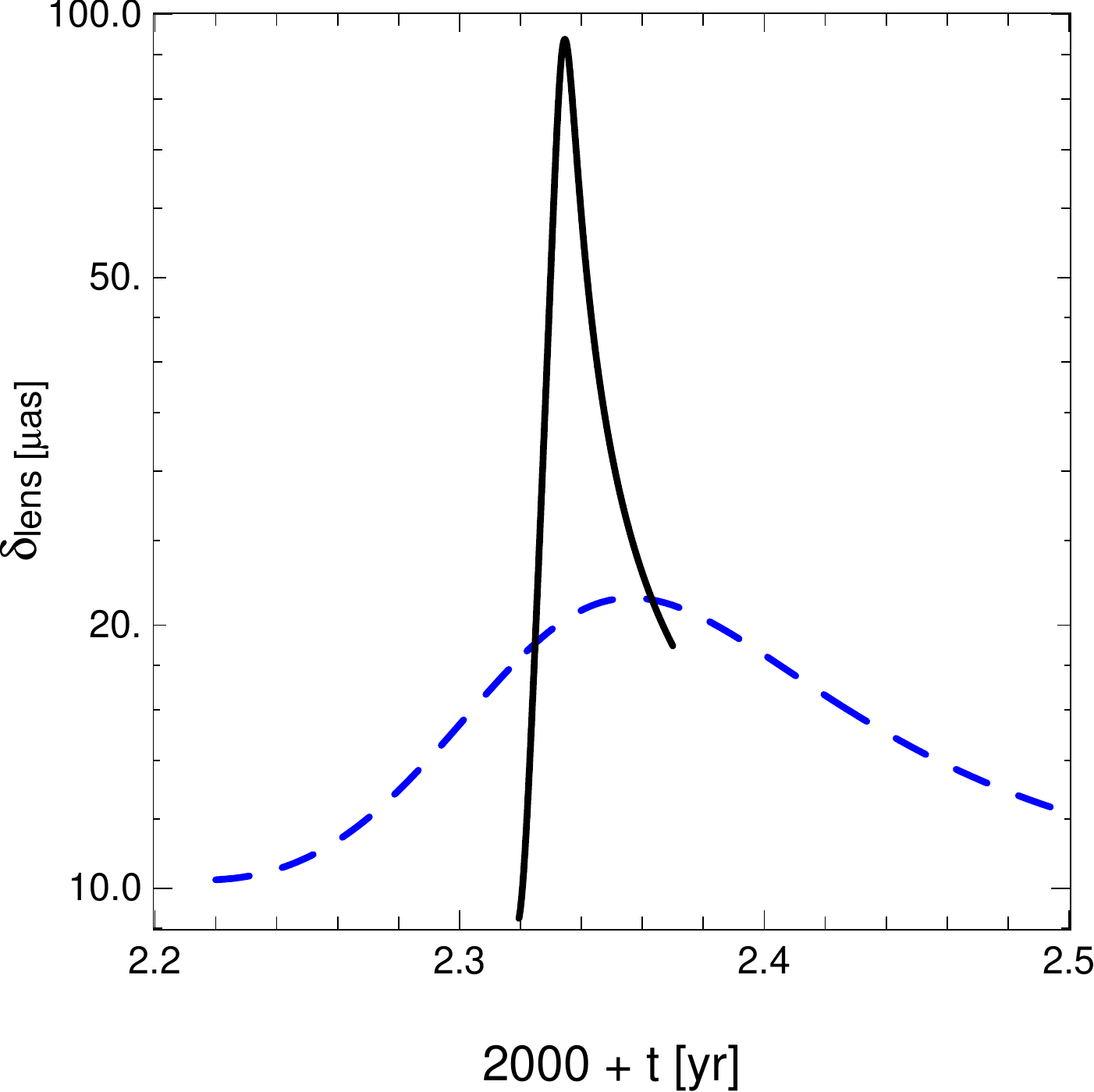} 
           \caption{Lensing effects affecting a portion of the S2 orbit (dashed blue line) and the E0 orbit (solid black). These simulated shifts in the astrometric observations are obtained considering a Keplerian model taking into account the lensing effects. This effect is added by using the analytical formulas developed in \cite{2006PhRvD..74l3009S}.}
   \label{fig:LE}
\end{figure}

To get an idea of the impact of neglecting lensing effects in stellar-orbit models on fitted orbital parameters, we decide to fit these parameters to astrometric and radial velocity data with this type of a model. The orbital parameters we fit are: the period of the orbit $T$, the semi-major axis $a_{\mathrm{sma}}$, the eccentricity $e$, the time of the pericenter passage $t_p$, the inclination $i$, the position angle of the line of nodes $\Omega$, and the angle from ascending node to pericenter $\omega$. To simulate the positions and radial velocities, we use a Keplerian
model and simulate the lensing effects using analytical formulae from \cite{2006PhRvD..74l3009S} (only applied when the star is behind the plane of sky containing Sgr~A*). Although not as accurate as a full-fledged numerical simulation, this simplistic model is sufficient for our purpose. No noise is added to the mock observations and we consider 1 000 dates to generate one complete orbit. 
We simulated two orbits: one based on the best-fit orbital parameters of the S2 star \citep{2009ApJ...707L.114G}, the other one for a fictional star (hereafter E0), which is closer to Sgr A* than S2. The parameters of each orbit are reported
in Table~\ref{table:unlensed_fit}. The impact of the lensing effects on the astrometric data is visible in Fig.~\ref{fig:LE}, where we only consider a portion of the orbits of the stars.

\begin{table*}[!t]
\begin{center}
\begin{tabular}{l|l|l|l|l|}
\cline{2-5}
& \multicolumn{2}{c}{S2} & \multicolumn{2}{|c|}{E0}\\[2pt]
\cline{2-5}
& \multicolumn{1}{c|}{Input} &  \multicolumn{1}{c|}{Fitted} &  \multicolumn{1}{c|}{Input} &  \multicolumn{1}{c|}{Fitted}\\[2pt]
\hline
\multicolumn{1}{|l|}{$T$ (yrs)} & 15.8 & $15.8 - 4 \times 10^{-6}$ & 1.5 & $1.5 + 10^{-6}$ \\[2pt]
\multicolumn{1}{|l|}{$a_{\mathrm{sma}}$ ($''$)} & 0.123 & $0.123 + 3 \times 10^{-7}$ & 0.0246 & $0.0246 + 8 \times 10^{-7}$ \\[2pt]
\multicolumn{1}{|l|}{$e$} & 0.88 & $0.88 + 2 \times 10^{-6}$ & 0.88 & $0.88 + 10^{-4}$ \\[2pt]
\multicolumn{1}{|l|}{$t_p$ (yrs)} & 2002.32 & $2002.32 - 2 \times 10^{-5}$ & 2002.32 & $2002.32 + 7 \times 10^{-6}$ \\[2pt]
\multicolumn{1}{|l|}{$i$ ($^\circ$)} & 135.25 & $135.25 + 3 \times10^{-4}$ & 100 & $100 + 9 \times 10^{-3}$ \\[2pt]
\multicolumn{1}{|l|}{$\Omega$ ($^\circ$)} & 225.39 & $225.39 + 2 \times 10^{-3}$ & 225.39 & $225.39 + 6 \times 10^{-3}$ \\[2pt]
\multicolumn{1}{|l|}{$\omega$ ($^\circ$)} & 63.56 & $63.56 + 10^{-3}$ & 10 & $10 + 3 \times 10^{-2}$ \\[2pt]
\hline
\end{tabular}
\end{center}
\caption{Input and fitted parameters for the S2 and E0 stars. The input parameters are those used to simulate the lensed mock observations. The fitted parameters are those estimated using an unlensed model.}
        \label{table:unlensed_fit}
\end{table*}


We then fit the mock data using a second Keplerian model, but neglecting the lensing effects. We finally re-computed mock (lensed) data using the fitted (non-lensed) parameters. The difference of the fitted parameters to the input parameters (also reported in Table~\ref{table:unlensed_fit}) seem small in relative value (the maximal relative difference is around $\simeq 10^{-5}$ for S2 and $\simeq 10^{-3}$ for E0), but
they are significant in the sense that using the fitted parameters instead of the input parameters in the mock data yields a significant error in the astrometry and in the radial velocity (up to 20~$\mu$as and 0.8~km/s respectively for S2, and 60~$\mu$as and 10~km/s, respectively, for E0).

The other point that should be discussed is the impact of using models neglecting lensing effects, for the detection of other effects affecting the orbit. To simulate these different effects, we consider the ray-tracing code GYOTO and the E0 star. We estimate the maximal astrometric influence of different effects such as Roemer, pericenter-advance, Shapiro and Lense-Thirring. We find that, in two orbital periods, the maximal effects reach: $\approx 220$~$\mu$as for Roemer, $\approx 240$~$\mu$as for the pericenter-advance, $\approx 90$~$\mu$as for gravitational lensing (see the solid black curve on Fig.~\ref{fig:LE}), and $\approx 8$$\mu$as for both Shapiro and Lense-Thirring. In two periods of monitoring, the Roemer and pericenter-advance effects dominate the lensing effects most of
the time. However, if we only consider the first pericenter passage, we note that the impact of lensing is of the same order of magnitude as the pericenter-advance and the Roemer effects. The gravitational lensing will thus probably interfere with detecting these effects. Beside, we find that higher-order effects such as Shapiro and Lense-Thirring are smaller than gravitational lensing. The detection of these smaller effects will be difficult if we do not take into account lensing effects in stellar-orbit models.

To summarize, we need to take into account the lensing effects in future stellar-orbit models. As shown here, it can not be neglected in most  cases. If these effects are not included in stellar-orbit models, they will probably prevent the measurement of other relativistic effects and we will not be able to efficiently constrain all key parameters, such as the spin of the black hole.


\section{Conclusion}
\label{sec:six}

The Galactic center is a unique laboratory to observe stars close enough to a compact object to test general relativity. Thanks to GRAVITY it will be possible to measure astrometric positions of stars orbiting Sgr A* with an expected astrometric precision of 10 $\mu$as. We have shown that GYOTO is extremely accurate even in complex configurations.
For the purpose of the interpretation of the future astrometric positions observed by GRAVITY, GYOTO is accurate enough to model star trajectories and fit the GRAVITY data. 

The various integrators implemented in GYOTO are all very accurate and fast. In most scenarios, the default parameters yield good results. On the other hand, when the astrometric error must be minimized or simply evaluated, or when computing time needs to be optimized, some effort must be put into choosing the optimal integrator and parameters. The scenario that we exhibit in this paper is particularly challenging: since each photon needs to be integrated over a long distance (several parsecs) before and after the interaction with the black hole, the deflection angle must be estimated to an extreme precision. One method to evaluate the numerical errors for a given integrator and parameters is to compare a typical geodesic that has been computed with these parameters to the same geodesic that has been computed with one of the most accurate setup: \texttt{Runge Kutta Fehlberg 78} and $\texttt{AbsTol}=\texttt{RelTol}~=~10^{-18}$.

In this paper, we also showed that lensing effects need to be taken into account to reach a model with an accuracy of 1 $\mu$as, even for stars located inside or in front of the plane of sky containing Sgr A*. The shift of the primary image can reach 5 $\mu$as for stars in this plane. The half angle of aperture, in which the astrometric shift is less than 1 $\mu$as, is about $50^{\circ}$. Out of this cone, the lensing effects must be taken into account to avoid a systematic error in each modeled astrometric position. It is especially true for the next generation of instruments such as MICADO on the E-ELT \citep{2010SPIE.7735E..2AD}, which are expected to observe stellar orbits with a high accuracy.

To constrain the nature of Sgr A* with future accurate instruments, we need to take into account the lensing effects in stellar-orbit models. Otherwise, fitting with models that neglect these effects will lead to inaccurate orbital parameter estimations and  prevent the measurement of others effects, such as the Lense-Thirring effect.


\section*{Acknowledgement}
This work was supported by the French ANR POLCA project (Processing of pOLychromatic interferometriC data for Astrophysics, ANR-10-BLAN-0511), and by the OPTICON project (EC FP7 grant agreement 312430) in "Image reconstruction in optical interferometry" WP4. We want to thank, Eric Gourgoulhon, Claire Somé, Jason Dexter and Stefan Gillessen for helpful discussions. We also would like to thank the referee and Fr\'ed\'eric Vincent for their advice, which enabled us to  improve the quality of the paper.

\appendix

\section{How to determine primary caustics and critical curves parameters with GYOTO?}
\label{app:A}

\begin{figure}[!h]
\begin{center}
        \includegraphics[scale=0.5]{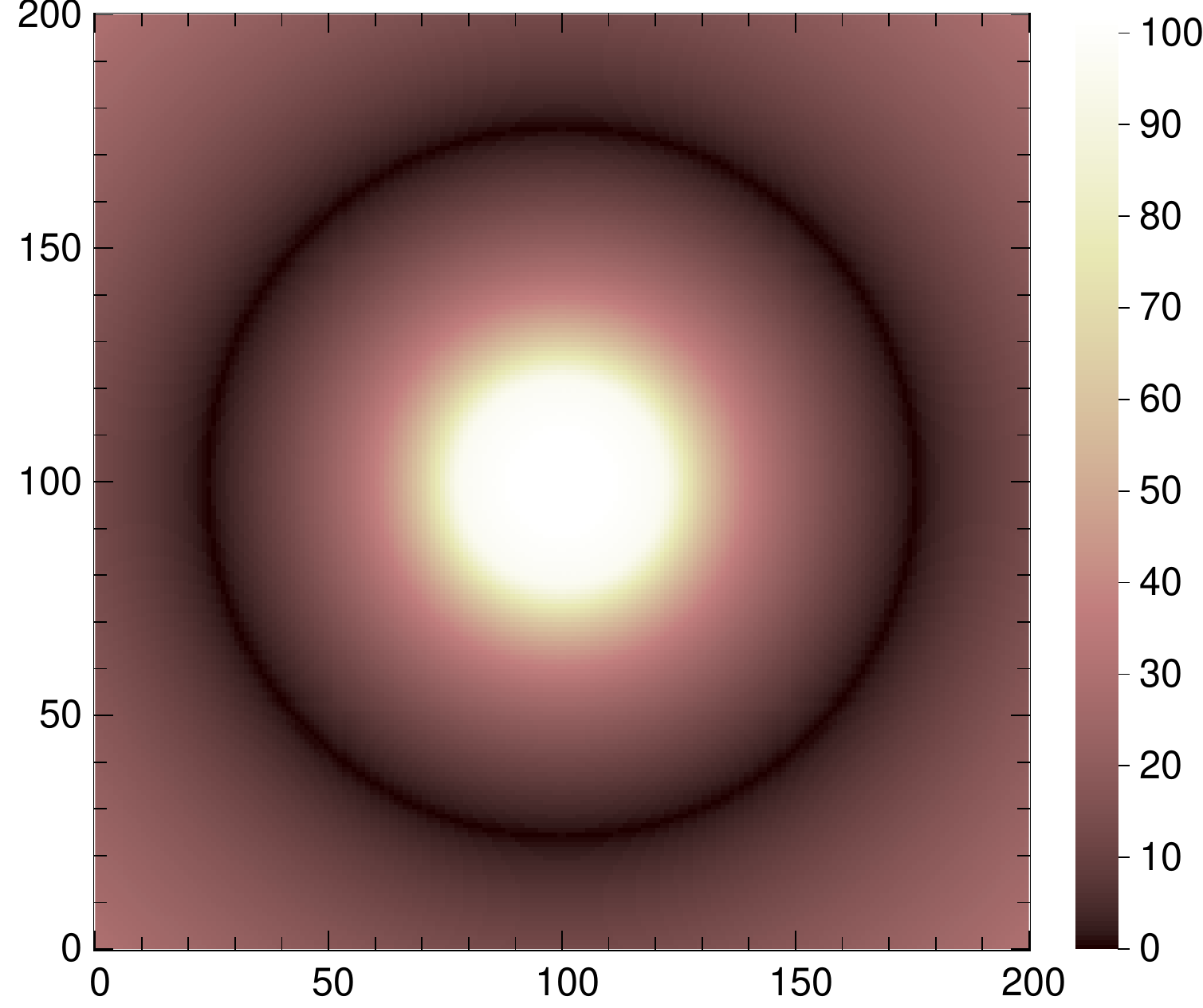}
        \caption{Illustration of the square root of the \texttt{MinDistance} function with a point source behind a Schwarzschild black hole. The black hole and the source are located at the center of the image. The source is at $100M$. The darker the pixels, the smaller the distance between the photon and the point source. The color-bar is labeled in $M$. The minimal and maximal values of the image is about $7 \times 10^{-4} M$ and $102 M$ (the maximal value of the color-bar is cut at $100 M$), respectively.}
                \label{fig:MD}
\end{center}
\end{figure}

GYOTO is a ray-tracing code integrating null geodesics backward in time. Each pixel on the observer's screen corresponds to the initial direction of each photon. When a photon reaches the star, the pixel illuminates and we get the star image, as presented in Fig. \ref{fig:ER2}.
However, we consider another type of image to estimate the three parameters $\Theta_E$, $\Delta,$ and $B_C$. We use a mathematical function named \texttt{MinDistance} in GYOTO. This represents the square minimum distance between the photon and the surface of the star. Zeroes of this function correspond to photons, which have reached the surface of the star. The advantage of this function is that it enables  images of the object to be located, even if no geodesic that is actually computed emanated from the star. Finding images of the object corresponds to finding minima of \texttt{MinDistance}, and checking that the minimum reaches zero.
To illustrate this function, we consider a point source behind a Schwarzschild black hole (see Fig. \ref{fig:MD}). 
The following subsections explain the measurement methods of the different parameters with the \texttt{MinDistance} function. 

\subsection{Angular position $B_{Cgyoto}$ of the caustic}
\subsubsection{Caustic point}
\label{app:A11}

Firstly, we are  interested in  measuring the angular position of caustic points. This means that we focus on caustics obtained in the Schwarzschild case or in the Kerr case, when the star is very far from the black hole. We note this angle $B_{Cp}$ in this subsection. To get the angular position of the caustic point, we need to estimate its position relative to the black hole, $y_{Cp}$. At this position, a critical curve is formed in the observer's sky. Thus, we have to search the position of the source $y_{Cp}$ leading to the formation of this critical curve.  If the source is not on the caustic point, minima of \texttt{MinDistance} will be located on the $\alpha$-axis and will correspond to the primary and secondary images. If the source lies on the caustic point, the two images merge and form the critical curve, and minima will be distributed all along the curve (see Fig. \ref{fig:MD}). Thus, to converge to $y_{Cp}$ we search the best minimum of \texttt{MinDistance} along the positive $\delta$-axis ($\alpha$=0) varying the position of the source $y_{s}$.  If we consider a Schwarzschild lens, it means that we look for the value of $y_{s}$, enabling us to reach the best minimum at the top of the critical curve. However, with a Kerr black hole, the critical curve is shifted from the $\delta$-axis (the center of the critical curve is no longer at $\alpha$=0) but we still search the best minimum of \texttt{MinDistance} along the positive $\delta$-axis. Even if we are not exactly on the axis passing through the top of the critical curve, the best minimum is reached when the critical curve is formed (because best minimums will be distributed all along the critical curve). 

The method we selected to find $y_{Cp}$ is a golden section search. This method is the same as the bisection method, but divides the interval by the golden number instead of 2. The convergence is faster than the bisection method.
The search process stops when the \texttt{MinDistance} function cannot be well estimated for two close $y_s$. The error on $y_{Cp}$ is given by the size of the last step used to divide the final interval. 
Using error propagation, we get the error on $B_{Cp}$ given by
\begin{equation}
\delta_{B_{Cp}} = \delta_{y_{Cp}} \left| \frac{1}{x_0 + |x_s| + y_{Cp}} \right|.
\end{equation}\\
The final error we consider is the biggest error evaluated for all distances of the source $x_s$ we selected.\\

The stop conditions and the method to estimate the maximal errors for the next parameters are the same as above. 

\subsubsection{Four cusps astroid}
\label{app:A12}

The first step to measure the angular position of the right cusp, noted $B_{Cr}$, is also to find its position relative to the black hole, $y_{Cr}$. To estimate $y_{Cr}$, we consider the primary caustics and critical curves properties that  we outlined in  Section~$\ref{ssec:KBHl}$. We use the link between the position of the source relative to the astroid, and images formed in the observer's sky (source outside or on the caustic lead to the formation of two images, source inside the caustic leads to the formation of four images). 
If we consider a source at the position ($x_s$,  $y_{Cr}$ , 0), we  observe two images in the observer plane, with one of these images forming on the left side of the critical curve (see the left triangle on the critical curve on the Fig. \ref{fig:C_CC}). The aim is to use a bisection method to converge to the right cusp.
Initially, we consider two different positions of the source: 
\begin{itemize}
        \item one inside the caustic, $y_{C_1}$. This is obtained using the same method presented in Section~\ref{app:A11}. We use a golden section search to look for the position of the source leading to the best minimum of the \texttt{MinDistance} function along the positive $\delta$-axis. In this case, it means that we search the position of the source leading to the formation of one image along the positive $\delta$-axis. If we take Fig.~\ref{fig:C_CC} into consideration, this image corresponds to the upper cross inside the critical curve.
        \item one outside the caustic, $y_{C_2}$. This is given by $y_{C_2} = y_{C_1} \pm \zeta$ with $\zeta$ superior to $\Delta_C/2$. The upper and the lower signs hold for the right cusp and for the left cusp, respectively. 
\end{itemize}
The next value of $y$ is given by the bisection method: $y_{C_3} = (y_{C_1}+y_{C_2})/2$. At each division of the interval we always satisfy the condition: one source outside (two images) and one source inside (four images). To estimate the maximal error of $B_{Cr}$ we also use the propagation error.


\subsection{Radius $\Theta_{Egyoto}$ and offset $\Delta_{gyoto}$ of the critical curve}
\label{app:A2}

These two parameters are estimated thanks to simple relations that depend on the right and the left radii of the critical curve relative to the black hole and measured along the $\alpha$ axis
\begin{equation}
        \Theta_{Egyoto} = \frac{r_l + r_r}{2},
\end{equation} 
\begin{equation}
        \Delta_{gyoto} = |\Theta_E - r_r|,
\end{equation} 
where $r_l$ and $r_r$ are the left and right radii, respectively.
In the Schwarzschild case the radii are equal.
To measure the right and the left radii of the critical curve, we need to find the position of the left and the right cusps, respectively. We perform the method explained before to obtain the two positions of the two cusps. Then, we estimate the radii using another golden section search. 
The final error on $\Theta_E$ and $\Delta$ is given by
\begin{equation}
\delta_{\Theta_{Egyoto}} = \delta_{\Delta_{gyoto}} = \frac{1}{2} (\delta_{r_l} +  \delta_{r_r}),
\end{equation} 
where $\delta_{r_l} = \delta_{r_{l1}} + \delta_{r_{l2}}$ and $\delta_{r_r} = \delta_{r_{r1}} + \delta_{r_{r2}}$ are the errors on the left and right radii, respectively. $\delta_{r_{l1}}$ and $\delta_{r_{r1}}$ are the errors obtained with the golden section search that was used to converge to the two radii. This is given by the last step of the interval. $\delta_{r_{l2}}$ and $\delta_{r_{r2}}$ correspond to the errors on the two radii owing to the errors made in the evaluation of the right $y_{Cr}$ and the left $y_{Cl}$ position of the two cusps, respectively. To evaluate $\delta_{r_{l2}}$ ($\delta_{r_{r2}}$), we estimate the left (right) radii obtained considering a source located at $y_{Cr} + \delta_{y_{Cr}}$ ($y_{Cl} + \delta_{y_{Cl}}$) and $y_{Cr} - \delta_{y_{Cr}}$ ($y_{Cl} - \delta_{y_{Cl}}$). We note the first radius $r_{l2_{+}}$ ($r_{r2_{+}}$) and the second radius $r_{l2_{-}}$ ($r_{r2_{-}}$).Thus, we get 
\begin{equation}
\delta_{r_{l2}} = |r_{l2_{+}} - r_{l2_{-}}|,
\end{equation} 
and
\begin{equation}
\delta_{r_{r2}} = |r_{r2_{+}} - r_{r2_{-}}|,
\end{equation} 
 
 For caustic points, we directly estimate the right and left radii at $y_{Cp}$. To estimate the errors, we just need to obtain the right and left radii for two values of $y_s$: $y_{Cp} + \delta_{y_{Cp}}$ and $y_{Cp} - \delta_{y_{Cp}}$, and use the same process explained in the previous paragraph.

\bibliographystyle{aa}
\bibliography{bbl}

\end{document}